%% file: 00_anchor.tex
\documentclass[sigconf]{acmart}
\AtBeginDocument{%
  \providecommand\BibTeX{{%
    \normalfont B\kern-0.5em{\scshape i\kern-0.25em b}\kern-0.8em\TeX}}}

\copyrightyear{2024}
\acmYear{2024}
\setcopyright{rightsretained}
\acmConference[IUI '24]{29th International Conference on Intelligent User Interfaces}{March 18--21, 2024}{Greenville, SC, USA}
\acmBooktitle{29th International Conference on Intelligent User Interfaces (IUI '24), March 18--21, 2024, Greenville, SC, USA}
\acmDOI{10.1145/3640543.3645214}
\acmISBN{979-8-4007-0508-3/24/03}

\usepackage{multirow}	
\usepackage{tabularx}	
\usepackage{xcolor}	
\usepackage{url}	
\usepackage{xspace}	
\usepackage{caption}	
\usepackage{subcaption}	
\usepackage{float}	
\usepackage{hyperref}	
\usepackage{cleveref}	
\usepackage{float}	
\usepackage{amsmath}
\hypersetup{	
    colorlinks=true,	
    linkcolor=blue,	
    filecolor=magenta,      	
    urlcolor=cyan,	
}	
\usepackage{enumitem}	
\setlist{leftmargin=6mm}

\begin{document}

\definecolor{Author1}{HTML}{e41a1c}  
\definecolor{Author2}{HTML}{377eb8}  
\definecolor{Author3}{HTML}{4daf4a}  
\definecolor{Author4}{HTML}{984ea3}  
\definecolor{Author5}{HTML}{ff7f00}  
\definecolor{Author6}{HTML}{fC9483}  
\definecolor{Issue1}{HTML}{e41a1c}  
\definecolor{Issue2}{HTML}{377eb8}  
\definecolor{Issue3}{HTML}{4daf4a}  
\definecolor{Issue4}{HTML}{984ea3}  
\definecolor{Issue5}{HTML}{ff7f00}  

\newif\ifsubmit	

\submittrue	

\ifsubmit

\newcommand{\One}[1]{#1}	
\newcommand{\OneA}[1]{#1}	
\newcommand{\OneB}[1]{#1}
\newcommand{\Two}[1]{#1}	
\newcommand{\TwoA}[1]{#1}	
\newcommand{\TwoB}[1]{#1}
\newcommand{\TwoC}[1]{#1}
\newcommand{\TwoX}[1]{#1}
\newcommand{\Three}[1]{#1}	
\newcommand{\Four}[1]{#1}
\newcommand{\FourX}[1]{#1}

\newcommand{\zinat}[1]{}	
\newcommand{\zinatIn}[1]{}	
\newcommand{\ray}[1]{}	
\newcommand{\rayIn}[1]{}	
\newcommand{\steve}[1]{}	
\newcommand{\steveIn}[1]{}	
\newcommand{\amanda}[1]{}	
\newcommand{\amandaIn}[1]{}	
\newcommand{\hemant}[1]{}	
\newcommand{\hemantIn}[1]{}	
\newcommand{\pq}[1]{}	
\newcommand{\pqIn}[1]{}	

\else
\newcommand{\zinat}[1]{\marginpar{\colorbox{Author1}{\textcolor{white}{ZA}} \textcolor{Author1}{#1}}}
\newcommand{\zinatIn}[1]{\colorbox{Author1}{\textcolor{white}{ZA}} \textcolor{Author1}{#1}}
\newcommand{\ray}[1]{\marginpar{\colorbox{Author2}{\textcolor{white}{RH}} \textcolor{Author2}{#1}}}
\newcommand{\rayIn}[1]{\colorbox{Author2}{\textcolor{white}{RH}} \textcolor{Author2}{#1}}
\newcommand{\steve}[1]{\marginpar{\colorbox{Author3}{\textcolor{white}{SP}} \textcolor{Author3}{#1}}}
\newcommand{\steveIn}[1]{\colorbox{Author3}{\textcolor{white}{SP}} \textcolor{Author3}{#1}}
\newcommand{\amanda}[1]{\marginpar{\colorbox{Author4}{\textcolor{white}{AH}} \textcolor{Author4}{#1}}}
\newcommand{\amandaIn}[1]{\colorbox{Author4}{\textcolor{white}{AH}} \textcolor{Author4}{#1}}
\newcommand{\hemant}[1]{\marginpar{\colorbox{Author5}{\textcolor{white}{HP}} \textcolor{Author5}{#1}}}
\newcommand{\hemantIn}[1]{\colorbox{Author5}{\textcolor{white}{HP}} \textcolor{Author5}{#1}}
\newcommand{\pq}[1]{\marginpar{\colorbox{Author6}{\textcolor{white}{PQ}} \textcolor{Author6}{#1}}}
\newcommand{\pqIn}[1]{\colorbox{Author6}{\textcolor{white}{PQ}} \textcolor{Author6}{#1}}

\newcommand{\One}[1]{\colorbox{Issue1}{\textcolor{white}{1:TheoreticalGround}} \textcolor{Issue1}{#1}}
\newcommand{\OneA}[1]{\colorbox{Issue1}{\textcolor{white}{1A:CiteGrpAnnotn}} \textcolor{Issue1}{#1}}
\newcommand{\OneB}[1]{\colorbox{Issue1}{\textcolor{white}{1B:CiteKnwlgGap}} \textcolor{Issue1}{#1}}

\newcommand{\Two}[1]{\colorbox{Issue2}{\textcolor{white}{2:MethodClarify}} \textcolor{Issue2}{#1}}
\newcommand{\TwoA}[1]{\colorbox{Issue2}{\textcolor{white}{2A:RenameExercise}} \textcolor{Issue2}{#1}}
\newcommand{\TwoB}[1]{\colorbox{Issue2}{\textcolor{white}{2B:Participant}} \textcolor{Issue2}{#1}}
\newcommand{\TwoC}[1]{\colorbox{Issue2}{\textcolor{white}{2C:Measures}} \textcolor{Issue2}{#1}}
\newcommand{\TwoX}[1]{\textcolor{Issue2}{#1}}

\newcommand{\Three}[1]{\colorbox{Issue3}{\textcolor{white}{3:Presentation}} \textcolor{Issue3}{#1}}

\newcommand{\Four}[1]{\colorbox{Issue4}{\textcolor{white}{4:StrengthenLimitation}} \textcolor{Issue4}{#1}}
\newcommand{\FourX}[1]{\textcolor{Issue4}{#1}}

\fi	

\newcommand{\BULLET}{\vspace{+.00in} \noindent $\bullet$ \hspace{+.00in}}
\newcommand{\accessdate}{09/13/2023}

\newcommand{\etc}{\emph{etc.}\xspace}
\newcommand{\ie}{\emph{i.e.,}\xspace}
\newcommand{\eg}{\emph{e.g.,}\xspace}
\newcommand{\etal}{\emph{et al.}\xspace}
\newcommand{\wrt}{\emph{w.r.t.}\xspace}
\newcommand{\aka}{\emph{a.k.a.}\xspace}

\title[Closing the Knowledge Gap in Data Annotation Interfaces]{Closing the Knowledge Gap in Designing Data Annotation Interfaces for AI-powered Disaster Management Analytic Systems}

\author{Zinat Ara\textsuperscript{\textdagger}, Hossein Salemi\textsuperscript{\textdagger}, Sungsoo Ray Hong\textsuperscript{\textdagger}, Yasas Senarath\textsuperscript{\textdagger}, \\ Steve Peterson\textsuperscript{\S}, Amanda Lee Hughes\textsuperscript{\textdaggerdbl}, and Hemant Purohit\textsuperscript{\textdagger}}
\affiliation{%
  \institution{\textsuperscript{\textdagger} George Mason University, Fairfax, VA \country{USA}\\ \textsuperscript{\S}Montgomery County Community Emergency Response Team, Gaithersburg, MD \country{USA}\\ \textsuperscript{\textdaggerdbl} Brigham Young University, Provo, UT \country{USA}}
  }

\renewcommand{\shortauthors}{Ara, et al.}

\begin{abstract}
Data annotation interfaces predominantly leverage ground truth labels to guide annotators toward accurate responses. 
With the growing adoption of Artificial Intelligence (AI) in domain-specific professional tasks, it has become increasingly important to help beginning annotators identify how their early-stage knowledge can lead to inaccurate answers, which in turn, helps to ensure quality annotations at scale.
To investigate this issue, we conducted a formative study involving eight individuals from the field of disaster management, each possessing varying levels of expertise.
The goal was to understand the prevalent factors contributing to disagreements among annotators when classifying Twitter messages related to disasters and to analyze their respective responses.
Our analysis identified two primary causes of disagreement between expert and beginner annotators: 1) a lack of contextual knowledge or uncertainty about the situation, and 2) the absence of visual or supplementary cues.
Based on these findings, we designed a Context interface, which generates aids that help beginners identify potential mistakes and provide the hidden context of the presented tweet.
The summative study compares Context design with two widely used designs in data annotation UI, Highlight and Reasoning-based interfaces.
We found significant differences between these designs in terms of attitudinal and behavioral data.
We conclude with implications for designing future interfaces aiming at closing the knowledge gap among annotators.
\end{abstract}




\maketitle
\input{01_intro.tex}
\input{02_related_work}

\input{03_S1}

\input{04_operationalizing_knowledge_gap}
\input{05_S2}
\input{06_implications_for_design}

\input{07_conclusion}

\begin{acks}
This research has been partially supported by The Office of Research, Innovation, and Economic Impact Fund (ORIEI) at George Mason University through grant \# 215135. The authors also wish to express their deep appreciation for volunteers of Montgomery County CERT, Maryland, USA, who provided valuable insights to understand better annotation interface designs throughout our design studies. 
\end{acks}


\bibliographystyle{ACM-Reference-Format}
\bibliography{99_ref}

\input{appendix}

\end{document}

%% file: 01_intro.tex
\section{Introduction}

As the quality of training sets directly contributes to the quality of Machine Learning (ML) models, research in data annotation focusing on producing high-quality labels has attracted attention from several communities, including HCI~\cite{choi2019aila,hong2020human, gao2021gnes}, Natural Language Processing (NLP)~\cite{bartolo2020beat}, Computer Vision~\cite{liu2019deep}, and beyond. 
Several data annotation approaches have focused on providing information cues that can help human annotators complete their work more efficiently, for example, by offering relevant examples~\cite{chung2019efficient} or highlighting words or images~\cite{choi2019aila}. 
Such efficiency-driven annotation designs generally seek to improve micro tasks that require less domain expertise, such as restaurant review sentiment classification~\cite{collab1} or object detection~\cite{madono2020efficient}.

As the goals of ML models become more contextualized and domain-specific~\cite{santos2019visus,yan2022flatmagic,ara2021ride, ara2021traffic}, however, many annotation task types require advanced knowledge that is beyond that of beginner annotators. 
Examples of such domain-specific problems include damage assessment~\cite{imran2022ai,cao2020building}, risk analysis~\cite{senarath2021mining}, information filtering for disaster management~\cite{pandey2022modeling}, medical image reading~\cite{lutnick2019integrated}, and many more~\cite{hong2019disseminating}. 
While less experienced are more available than those with extensive domain experience, helping the less experienced to annotate ``like a pro'' by solely relying on efficiency-driven designs can impose challenges.
Prior research has emphasized the significance of incorporating a diverse range of knowledge and perspectives~\cite{austin2003transactive,10.1145/3544548.3580645, RUSSO2021117695}, highlighting the need to understand the implications of the knowledge gap between beginners and experts in modern data annotation user interface design. 
However, few approaches have been explored the way to close the knowledge gap in designing data annotation user interfaces. 


In this work, we seek to understand how the knowledge gap can impact the annotation performance of beginners and characterize common reasons that lead beginners to make different decisions than experts.
We further seek to explore how annotation design strategies can provide information that mitigates the knowledge gap. 
To understand and measure this knowledge gap, we conducted a formative study (S1) with a diverse group of eight participants (Experts and Beginners) in different sub-domains of disaster management. 
We provided distinct sets of one thousand tweets related to the disaster Hurricane Ian~\cite{Ian} to both groups and requested them to annotate whether each tweet pertained to any emergency event involving transportation means, damaged infrastructure, or was unrelated to the context. 
Next, we interviewed the annotators to understand the rationale behind their differing annotation decisions when compared to expert annotators. 
Our analysis revealed two common reasons for these disparities: \textbf{confusing words}: messages that contain related words to transportation and infrastructure, such as ``cars'' or ``bridges'', but the ground truth is negative and \textbf{hidden context}: instances where messages lacked readily apparent related words but required an ``institutional level of insights'' to interpret positive relevancy. 

In our summative stage (\textbf{S2}), we aimed to evaluate how an AI-assistive design built based on S1 insights can impact data annotation task performance compared to widely used state-of-the-art designs. 
In S2, we built the \textbf{Context} interface which provides cues to detect confusing words commonly misinterpreted by the annotators and reveal the hidden context derived from the annotation differences between domain experts and beginners that we identified in S1. 
We built two additional designs that represent commonly employed annotation user interface designs. 
One of the designs is the \textbf{Highlight} interface that color-codes relevant words~\cite{choi2019aila, gooding2023study}. 
The other design is the \textbf{Reasoning} interface that explains ``why'' the message can be interpreted as positive or negative by using the advanced Large Language Models (LLMs)~\cite{chatgpt, bai2022training}. 
Using the 3 conditions, we conducted an experimental study with 13 Community Emergency Response Team (CERT) volunteers who are beginners in their domains. 
Across the three conditions, we measured how their behavioral and attitudinal annotation task performance varied. 
We observed that the Highlight and Context interfaces exhibited nearly identical behavioral accuracy. 
The Highlight interface proved to be the fastest among the three designs. 
In terms of attitudinal performance metrics, the Context design outperformed the other two, leading annotators to perceive it as the most effective in reducing the knowledge gap.

This work offers the following contributions:
\begin{itemize}
    \item \textbf{Empirical understanding of the knowledge gap in data annotation:} Through S1, we offer empirical insights into the knowledge gap that exists between data annotators, specifically within the disaster management domain. We accomplish this by eliciting annotations from two groups of beginner and experienced annotators. Through interviews with beginner annotators, we identify the two common reasons behind annotation differences between beginners and experts—confusing words and hidden context.
    \item \textbf{Design for mitigating knowledge gap}: 
    Based on the S1 findings, we develop a novel interface design that leverages the two common reasons behind annotation differences between beginners and experts. 
    \item \textbf{Experimental study:}  To understand how the new design can affect beginners' annotation mitigating the knowledge gap, we conducted S2 and reported results. 
    \item \textbf{Implications for design:} Based on S1 and S2 findings, we discuss how annotation interfaces can leverage our insights to help beginners reduce the knowledge gap in the future.
\end{itemize}

%% file: 02_related_work.tex
\section{Related Work}
In this review, we describe how research in group work has been applied to design annotation interfaces. 
We then examine existing data annotation interface studies. 
By synthesizing insights from both review directions, we conclude 
with the necessity of more deeply understanding knowledge gap-driven design in annotation interface research.

All forms of collaboration stem from a similar motivation; group work can be productive because it allows individuals to share and gain insights they might not have discovered independently~\cite{cabrera2006determinants}.
By pooling the knowledge and expertise of all members, group work can be executed more efficiently and accurately~\cite{austin2003transactive, hong2018collaborative, hong2019design, 10.1145/3290605.3300522}. 
Several studies in HCI, especially Computer-Supported Cooperative Work (CSCW), have built interactive systems and designs that apply in group work settings, such as collaborative searching, collaborative information seeking, and beyond~\cite{shahi2021amused, serrano2013cloud, hong2019design, amershi2008cosearch, bentley2017searchmessenger}. 
In the research on data annotation interfaces, a line of studies apply the spirit of group work--``the whole is greater than the sum of its parts''--especially when the ground truth is not deterministic and individual viewpoints can matter.
For instance, Chang et al.~\cite{10.1145/3025453.3026044} introduced a tool ``Revolt'' facilitating group decision-making for collaborative crowdsourced workers. 
They leveraged the disagreement between the annotators to identify ambiguous concepts and provide more options for decision-making based on the disagreement feedback. 
Upchurch et al. \cite{Upchurch_Sedra_Mullen_Hirsh_Bala_2016} present an online game allowing annotators to refine the majority vote labels to build consensus among them while labeling image data. 
\One{\OneA{Numerous research works have been focused on group decision-making that theorizes how a group, either multiple humans or humans and AI- can explain different perspectives and resolve potential disagreements.
For example, Sutcliffe introduced a Small Group Theory for designing CSCW systems that employ model-based analysis of group members, technology support, and design principles derived from the theory~\cite{10.1145/1082983.1083119}. 
Hong et al. investigated the impact of single-user designs on consensus-building processes~\cite{10.1145/3359208} and, to foster group consensus, they introduced Collaborative Dynamic Queries~\cite{collab1}. 
This approach allows a group to filter decision criteria while sharing the choices made by others. 
Similarly, Kairam and Heer~\cite{10.1145/2818048.2820016} illustrated how crowd-parting analysis at the intermediate level offers insights into sources of disagreement not readily apparent when examining individual annotation sets or aggregated results. 
Brachman et al. suggested a system where AI assists in identifying cases where the majority vote by labelers is incorrect, employing automation for conflict resolution tasks~\cite{10.1145/3555212}. 
The outcome demonstrated that automated conflict resolution enhances user accuracy and efficiency.}}

Past studies have developed intuitive interfaces that simplify the task of labeling, incorporating features like drag-and-drop~\cite{shneiderman2000direct} or batch labeling~\cite{ashktorab2021ai}, highlighting~\cite{choi2019aila}, and leveraging language-based models~\cite{bai2022training}.
Stureborg et al. grouped more similar contents together and different kinds of pass logic to coordinate between the crowdsourced annotators in multi-labeling tasks~\cite{Stureborg_2023}.
Gooding et al. performed a comparative study on annotation interfaces for summarization tasks trained in-house annotators with backgrounds of different proficiency levels~\cite{gooding2023study}.
\OneB{The majority of tasks discussed in the earlier literature fall into the category of "microtasks," which can be performed by almost anyone.
In contrast, we define our target annotation task as one that demands advanced knowledge or experience for making accurate decisions, distinguishing it from "microtasks." 
Our task is particularly concerned with the knowledge gap between beginners and experts, aligning with the focus presented by Wilkins et al~\cite{10.1145/3512940}. 
In their qualitative study, Wilkins et al. investigated how 38 knowledge workers (21 freelancers and 17 employees) apply knowledge, demonstrate skill in their work, and mobilize resources to address knowledge needs—a phenomenon referred to as the "knowledge gap." 
When designing technology to alleviate this knowledge gap, it is crucial to address the social-technical gap, as this represents a central challenge for the CSCW~\cite{ackerman2000intellectual}.}
Several studies have highlighted how this knowledge gap can generate unwanted productivity erosion. 
For instance, Kapania et al. studied that when annotators lack access to essential information, clear guidance, and opportunities for collaboration, a knowledge gap can emerge among them~\cite{10.1145/3544548.3580645}. 
Expert annotators possess a deep understanding of the domain and context, enabling them to provide more accurate and consistent annotations whereas beginners may lack this critical expertise, leading to errors and misinterpretations in their annotations \cite{Robson_Searston_Edmond_McCarthy_Tangen_2020, wang2022ai, RUSSO2021117695}. 
Annotation tasks are designed to establish a definitive ``ground truth'' label for training machine learning models to minimize the influence of annotators' knowledge gap~\cite{10.1145/3544548.3580645}. 

Through the literature review, we identified that the insights from group work have been mostly applied to subjective tasks with no deterministic ground truth rather than objective tasks.
In handling the objective tasks, meanwhile, we also identified that the research in data annotation design has put more emphasis on supporting microtasks where domain-specific expertise is less important. 
Based on the review results, we were motivated to further understand how the knowledge gap can manifest in objective tasks, what are the common reasons, and how new data annotation user interface designs can effectively support ``beginner'' annotators.

%% file: 03_S1.tex
\section{Formative study (S1)}
Our formative study's objective is to analyze the knowledge gap between experienced and beginning annotators while they identified transportation-related 
events on social media during a disaster. 
S1's Research Questions (RQs) are as follows:
\begin{itemize}
    \item \textbf{RQ1.} How does the variance in expertise between Beginners and Experts introduce differences in conducting annotations?
    \item \textbf{RQ2.} What are the typical reasons when an annotator's knowledge gap can incur disagreement in annotation tasks?
\end{itemize}

\subsection{Recruitment}
For S1 we recruited 
eight participants 
in different sub-domains of disaster management with varying levels of expertise. 
One of the authors with expertise in the field reached out to the participants via email to inquire about their interest in participating in the study. 
From the interested participants, we selected four Experts that included an EMT specialist and transportation manager from the county-level government, along with two emergency managers and a cybersecurity analyst from the federal-level government. 
The four remaining participants were community members affiliated with a Community Emergency Response Team (CERT) at the county-level of government, categorized as Beginners for the purposes of this study. 
CERTs in the United States offer standardized training and organization, serving as reliable resources for disaster response entities like transportation agencies~\cite{cert}. 
When the operation is activated, CERT volunteers can assist formal humanitarian organizations across different disaster response and management tasks, expanding their roles to virtual support, such as social media analysis. 
Each participant was compensated with a gift card for their participation in the study.

\subsection{Method}

To gain a deeper understanding of current approaches to extract relevant information during disaster 
events, 
the common challenges they encounter, and how the knowledge gap can impact their workflow, we conducted open-ended interviews with Expert group participants (P4, P5, P7, P8). 
These interviews were conducted remotely through the Zoom platform and lasted approximately one hour each. 
Prior to commencing the interviews, we obtained informed consent from all participants. 
Interview questions centered around the following key themes: 1) the expert’s overall workflow for identifying significant events on online intelligence systems, 2) the procedures employed for training less experienced volunteers and the corresponding challenges, 3) the impact of the "knowledge gap" between less experienced and experts when analyzing online intelligence, and 4) the desired features sought to speed Beginners' experience acquisition in this domain.

The next phase was the ground truth collection phase.  
Creating a gold standard is a fundamental requirement for performance evaluation and accuracy verification~\cite{10.1080/01431160802672864}. 
The purpose behind collecting ground truth data is to pinpoint crucial instances where annotators exhibit the most disagreement and determine their correct labels. 
These datasets would then be employed during the design evaluation phase, using various interface designs, to assess whether beginners can annotate them with improved accuracy. 
As part of this phase's requirements, we conducted interviews with four Beginner CERT volunteers (P1, P2, P3, P6) to gain deeper insights into their decision-making processes. 
\TwoB{We analyzed the disagreement data with the help of a transportation expert (exclusive from our recruited expert participants) to prepare our test datasets.}
This analysis also guided us in identifying design considerations for an annotation interface that might reduce the knowledge gap and enable Beginners to make more accurate decisions.
\TwoB{ To summarize, we split our participants into (Experts: P4, P5, P7, P8), (Beginners: P1, P2, P3, P6), and one transportation expert analyst for verifying ground truth data.}

\subsection{Understanding Problem Scope}

\TwoB{This stage of the study is focused on addressing our initial research inquiry, denoted as RQ1.
Recruited expert participants have field experience in disaster management and have witnessed first-hand the difference between experienced and inexperienced personnel.
Their practical insights are instrumental and relevant in qualitatively exploring knowledge gaps within this domain.
The Experts also shared their experiences with existing supervised machine learning systems used in their practice and deliberated on applicable considerations for these systems.}
These systems have undergone training to recognize disaster-related information aligned with specific disaster mission objectives.
P5 emphasized the importance of considering various sources in these systems, such as social media platforms, television, or the Internet, for gathering information. 
All of these sources can provide different kinds of information from different populations that together can form a more complete picture of emergency situational awareness. 
However, one prominent challenge that emerged from our discussions pertained to the contemporary deluge of data and the difficulty of assessing its accuracy. P5 elaborates, \textit{``The volume of data and the lack of awareness as to its legitimacy. Certainly, misinformation can lead us down a path to being challenging to have to validate and then ultimately throw out the information as far as being relevant''}.
Incorrect labeling and misinformation compound this challenge, further complicating efforts to verify the credibility of information within these systems.

Annotation agreement and feedback on that is also important in this context. 
P5 remarked from his past annotation task experience that there 
were not 
many differences or mistakes between experts' and beginners' decisions, rather among the potential reasons for making those decisions, \textit{``There were not that many errors or mistakes made by the less experienced versus experienced annotators. I just think the difference between potentially one of the reasons there''}. 
Another concern is the challenge of accurately assessing the severity of a situation based solely on media, particularly when it comes to events like natural disasters. 
For example, P4 mentioned, \textit{``People post things about a situation, it may sound terrible, but it may be very localized. 
There may be some flooding in the county or in a specific location. It is not like there is flooding all over the county''}. 
P7 underscores the importance of considering context and perspective when assessing the severity of a situation, for instance, P7 said, \textit{``My car is in high water, and I can not cross the road. Well, it is bad for you, but I do not know if that is an emergency for others''}. 
This denotes the subjectivity of emergencies urgent situations for one person may not be applicable to others. 

P7 emphasizes the idea that people's perspectives on whether something is good or bad can vary significantly based on their position and experience. 
As P7 says, \textit{``Your reason for thinking this is bad or good is very different, based on kind of the position you are in. If a junior person says to me, it is really bad for them because of this reason. And then I say, I do not think it is so bad, because, from a business perspective, this is not what we cover''}. 
The statement exhibits how one's professional background, experience, and responsibilities can shape their perception of a situation. 
Additionally, it suggests that more experienced individuals may tend to have a broader perspective and not view every issue as critically as less experienced individuals might.

\subsection{Ground Truth Collection}

Ground truth collection was conducted in three steps, 1) Collecting samples from social media, 2) Performing annotation task, and 3) Disagreement Analysis.

\subsubsection{\textbf{Collecting samples from social media:}}
\label{sample}
In the process of data collection, our primary focus was on the recent disaster event Hurricane IAN~\cite{Ian}. 
We gathered data (publicly posted status and text messages) from the Twitter social media platform~\cite{twitter} specifically within the timeframe of September 23, 2022, to October 2, 2022, covering both the event itself and the aftermath.

For the annotation task, two classes were defined: Transportation Means (TM) and Damaged Infrastructure (DI). 
TM was defined as the means used to move people and/or goods from one place to another and must have operational value to public safety mission, for example, transportation officials may call for a debris management team to go and remove the inoperable vehicle from the roadway during the disaster or emergency events. 
DI was defined as foundational structures and systems for transporting people and goods that have been partially or completely damaged.


We applied an existing methodological framework ~\cite{peterson2019} that describes the process of collecting relevant data for disaster management agencies. 
The process included the use of domain expert-provided keywords 
to search/filter 
transportation-specific 
messages from 
Twitter.  
 ChatGPT~\cite{chatgpt} was used to enhance the preliminary 
set of keywords. Given Hurricane Ian made landfall in the Southwest District of Florida we queried the tool based on common transportation means found in that geographical region. 
Additionally, we queried the names of bridges, causeways, ports, highways, bus and rail services, and county-level transportation agencies in Lee, Charlotte, etc. 
A total of 475 keywords were identified as relevant to the context and verified by a domain expert. 
Using these keywords, we filtered out and curated tweets, resulting in an updated final sample of 4,000 data points for our initial annotation task. 

\subsubsection{\textbf{\Two{Performing annotation task:}}}
\label{annotation}

\TwoX{From the pool of 4,000 tweets gathered in section~\ref{sample}, we divided them into four distinct datasets and assigned each dataset to an individual annotator. 
Each annotator was tasked with labeling a set of 1,000 tweets that were exclusive to the other annotators. 
The annotation involved categorizing each tweet text as either TM, DI, or both. 
Participants also had the option to designate tweets as IR (Irrelevant) to the context. 
The primary aim of this task is to gather annotations from beginners, which will later be cross-verified by experts to identify any disagreements (labels differing from the expert's decision).
A similar interface to ~\cite{senarath2022citizen} was used to execute this task. 
The data collection process occurred asynchronously, spanning a total duration of one week.}

\subsubsection{\textbf{Disagreement Analysis:}}

\Two{This stage of the study focuses on addressing our second research inquiry, RQ2.
In this step, we analyze the collected instances of the annotation set from~\ref{annotation}. 
A domain expert with decades of experience in the emergency management profession conducted a comprehensive 
analysis of these tweets. 
The Expert reviewed the instances and determined which label was correct for a single tweet. 
If the expert's decision differs from the annotator's, then disagreement occurs.
 
The result revealed that out of 909 tweets where disagreements occurred, approximately 51\% (459 cases) were accurately labeled by the Expert, approximately 43\% (392 cases) were accurately labeled by Beginners, and 58 cases (approximately 6\%) remained inconclusive, with neither party making the correct decision.}
Furthermore, these disagreements were categorized into distinct themes based on their types, such as classification errors, institutional insights, lack of visual cues, language barriers, and matters of opinion.

We organized online sessions with Beginners to collect insights about their decision-making processes. 
\Two{Each session incorporated two distinct types of exercises. 
Both of these approaches collect natural user behavior in a relatively unobtrusive manner over an extended period, providing some insights into individuals' thought processes as they engage in activities. 
They are particularly valuable for comprehending the underlying reasons behind tasks that require focused attention~\cite{Russell2014LookingBR}.}
\begin{enumerate}
\item \TwoA{Data annotation session}: In this task, we reviewed 50 samples drawn from the pair-wise annotation project, specifically targeting instances where Beginners made incorrect decisions. Our selection process prioritized cases falling under the "Institutional insights" category due to hidden contextual indicators suggesting operational relevance, requiring annotators with disaster management expertise. We also included samples from other disagreement categories. Our primary focus was uncovering two key insights for each tweet: the rationale behind the prior class selection (TM, DI, or IR) in annotations and potential alternative interpretations. We recorded participants' responses, contributing to a comprehensive analysis.

\item \TwoA{Follow-up retrospective interview}: This session encouraged annotators to group and articulate the reasons behind their decisions. 
We asked each participant to list 5 common reasons for grouping their annotation decisions. 
Among the most frequently identified reasons were: a lack of contextual knowledge or uncertainty about the situation (mentioned by 6 participants), the absence of visual and supplementary cues (mentioned by 5 participants), consideration beyond mere keywords, the presence of low-quality tweets, relevance to transportation or infrastructure, clear indicators of events, and the inclusion of topics unrelated to the emergency situations.
\end{enumerate}

\subsection{Design Considerations}
After analyzing feedback from both expert and beginner annotators, it became evident that annotating social media data within the context of disaster management presents a multifaceted challenge~\cite{hughespeterson2014}. 
Annotators frequently encounter difficulties in making accurate decisions due to the nuanced nature of interpreting messages in this domain, a process heavily influenced by their individual levels of expertise. 
Individuals with more experience tend to possess a deeper understanding of both knowledge and context within a given field, as P7 mentioned, \textit{``Experience tends to know more about the knowledge and of the context and it is because of their experience. They can like to take those events from their own experience and it is easy for them''}. 
We have pinpointed key challenges and explored potential design strategies aimed at bridging the knowledge gap, enabling beginners to make decisions equivalent to those of experts in this complex task.

\subsubsection{\textbf{Revealing hidden message context}}
The most frequently cited factor influencing annotation decisions is an awareness of the message's context. 
Less experienced annotators often struggle with grasping the correct context of tweets, particularly when it relates to the field of disaster 
management. 
Having more context or information about a situation can simplify the process of deciding what class or category to assign, as P1 mentioned, \textit{``If you see more context, it makes it easier to make a decision as to what the label would be''}.  
Providing users with easily accessible definitions and customized resources could enhance their understanding of unfamiliar terms and contexts, ultimately improving their ability to complete tasks or missions effectively. 
For example, P5 said, \textit{``It could be beneficial for less experienced users to understand and comprehend terms and contexts a little bit better where they could have sort of like a dictionary that gives the definitions of what their mission is or a more customized resources pointing people to getting more information about the tweet''}. 
P6 also discussed about similar issue of insufficient context for a particular topic, due to a lack of visual cues or inaccessible links. 
If annotators had access to these cues, it might provide the missing context. 

\subsubsection{\textbf{Providing insights from past annotation decisions}}

Exploring different strategies for better grasping the importance of elements in the annotation task, possibly by relying on the collective wisdom or opinions of others, has been useful in past literature~\cite{10.1145/3025453.3026044, Chen_2021}. 
P4 mentions sharing what other people thought or getting insights from others' perspectives might be more helpful in determining the significance of certain aspects of the task, as he remarked, \textit{``This was relevant because of this aspect or something like that. So maybe if you pointed out what other people thought, would be helpful as well''}. 
Annotation decisions by experts also provide clear and well-illustrated ideas of how the annotation task should be performed. 
Beginners can learn from these examples by observing how experts approach complex or ambiguous cases. 
Some participants also talked about example-based guidance in a learning context and role-playing exercises can be helpful for teaching as they provide practical, real-life scenarios for learners to engage with. 
For instance, P7 said, \textit{``Example-based guidance, such as what can be the good things, what can be the bad things that you have to kind of role-playing exercise would be super beneficial''}.

\subsubsection{\textbf{Emphasizing on class-relevance hints}}

The participants also discussed the advantages of emphasizing significant keywords or phrases associated with a specific class, a practice that has been utilized in previous research as well~\cite{choi2019aila, gooding2023study}. 
By providing a visual aid that draws attention to the most critical elements within the text or content being annotated, such as highlighting key terms or phrases, annotators can quickly locate and focus on the information that is directly relevant to the task. 
As P1 remarked, \textit{``Trucks, commercials or keywords related to transportation means, or any kind of damaged infrastructure will be actually helpful to you to make these decisions''}. 
Emphasizing the important information can help annotators focus on a particular situation where they see significant damage resulting from an event or incident. 
For example, P2 said he was specifically focused on extreme cases, \textit{``I guess I was looking at it as the extreme damage was done and where our emergency resources are needed''}. 
While highlighting keywords can be a valuable starting point for annotators, it's essential to consider the broader context surrounding those words to grasp their intended meaning fully. 
For instance, P4 said, \textit{``Highlighting the relevant keywords and if someone is instructed paying attention to anything but where it mentions driving or your car, then some of those words might get picked which really have nothing to do with the disaster event''}.

\subsubsection{\textbf{Providing clues for potential errors}}

Making informed choices during the annotation task is crucial in the context of disaster management. 
We observed that some annotators prioritized a set of keywords over the contextual information because these keywords appeared significantly pertinent to a specific class, even when the tweet itself did not pertain to an emergency situation. 
This can occur due to the complexity of certain tweets, which makes it challenging to correctly categorize them, as well as differences in annotators' experience levels. 
For example, P3 mentioned, \textit{``It mentioned car, that’s why I chose TM,  but this one probably should have been marked irrelevant because it doesn't mention anything about the actual storm. At 70\% time, I was probably selecting classes based on keywords, and probably for 30\%, filtering it of what could be most useful for them''}. 
While false positive cases may not significantly impact operations, false negatives for transportation sector in 
emergency scenarios can have a substantial adverse effect on the situation. 
For instance, P8 said, \textit{``If we make a mistake in the wrong place at the wrong time, we could actually get somebody killed''}. Providing hints for potential errors or ambiguous classifications is immensely helpful to annotators as it can offer guidance and clarification in situations where there might be uncertainty or complexity. 
When annotators encounter ambiguous cases having those kinds of hints can help them make more informed decisions.

\subsubsection{\textbf{Elaborating message content depending on class}}

Explaining possible reasonings for a message belonging to a particular class can also be helpful to annotators, as users can understand better why a message fits into a specific category or why it does not. 
This enhances their decision-making process. 
To support this P5 remarked, \textit{``If people were trying to understand the relevancy of transportation means and damaged infrastructure, having a system where that could provide some local knowledge within it would be tremendously beneficial''}. 
This highlights the importance of local insights and expertise in understanding the practical implications and significance of transportation and infrastructure issues.

%% file: 04_operationalizing_knowledge_gap.tex
\section{Annotation interface design}


Based on the design considerations, we developed three annotation interfaces for providing aids to the annotators: 1) Highlight, 2) Reasoning, and 3) Context. 
The Highlight and Reasoning interfaces are designed based on state-of-the-art techniques used in data annotation~\cite{choi2019aila,gooding2023study,bai2022training}. 
Since our S1 revealed significant challenges in the decision-making process, we introduced the Context interface, which incorporates two types of hints: 1) highlighting potentially confusing words and 2) hidden context within the presented tweet.

\subsection{\textbf{Common features}}

All interfaces have been created to facilitate text annotation~\cite{senarath2022citizen} by presenting individual Twitter messages one at a time and inquiring about their relevance to specific classes. 
Each question is accompanied by two radio button options (``yes'' and ``no'') inquiring whether the displayed tweet is associated with that specific class as shown in Figure \ref{fig:fig_prototypes}. 
The "Confirm and submit" button will be disabled by default when no option is selected. 
After the users choose an option they can proceed to the subsequent question by clicking the "Confirm and submit" button. 
To maintain the fidelity of our experimental conditions, questions for each tweet are initiated per the tweet's ground truth label (considering both false positive and false negative cases). 
Further elaboration on the sampling method details can be found in Section~\ref{datasample}. 
Users can also track the number of completed and remaining tweets from a progress status bar.
The annotation task is executed for two classes (TM and DI). 
If both class options are labeled as "no," we categorize the tweet as irrelevant (IR) for the given context.

\subsection{\textbf{Highlight}}

\begin{figure*}
  \centering
  \includegraphics[width=0.85\textwidth]{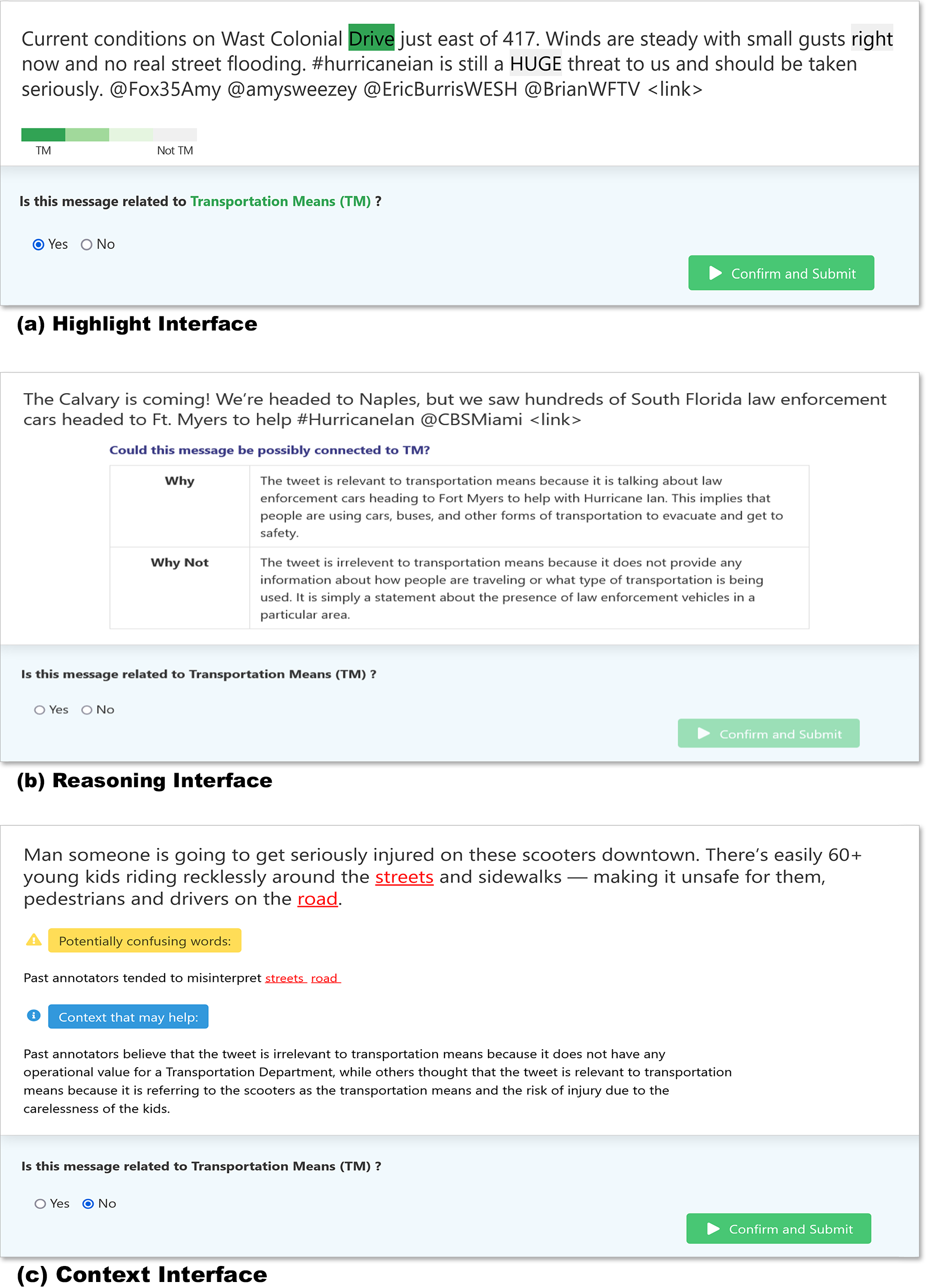}
  \caption{Providing assistance and inquiring if the tweet belongs to TM class in (a) Highlight interface, (b) Reasoning interface, (c) Context interface}
  \vspace{0in}
  \label{fig:fig_prototypes}
  \Description{Providing assistance and inquiring if the tweet belongs to TM class in a) Highlight interface, b) Reasoning interface, c) Context interface}
\end{figure*}

The Highlight interface employs color-coded schemes to highlight specific keywords within a single tweet. 
We highlight relevant and irrelevant words (tokens) of the text to help the annotator pay more attention to the tokens that can potentially be indicative of the correct label. 
Moreover, we consider different intensities in highlighting the tokens to represent how much a token is relevant or irrelevant to the label. 
For example, the token ``Drive'' with darker shades indicates a stronger connection (TM), while the lightest shaded tokens ``right'', and ``HUGE'' represent the weakest association (Not TM), as shown in Figure \ref{fig:fig_prototypes} (a).


In our formative study, we used 4000 samples to find disagreement cases. 
We use samples from the agreement part to extract relevant and irrelevant tokens to each label and calculate their relevancy measures, as follows:
\begin{enumerate}
    \item For each sample, we create a list of candidate tokens. These tokens include nouns, verbs, and named entities that exist in the sample. Here, we use Spacy (an open-source library for NLP) to extract these tokens. 
    \item To calculate the relevancy of each token to a class $C$, we employ normalized Pointwise Mutual Information (nPMI) measure \cite{bouma2009normalized}. First, we calculate: \\ $PMI(t,C) = \log (p(t,C)/(p(t)p(C)))$, where $p(t,C)$ is the probability of a sample containing token $t$ and is annotated as class $C$, $p(t)$ is the probability of a message containing token $t$, and $p(C)$ is the probability of a message being annotated as class $C$. Then, we use $nPMI(t,C) = PMI(t,C)/(-\log_{2} p(t,C))$ to normalize PMI measure and map the PMI measure to the range of [-1,1]. This measure is used to measure the relevancy of the token $t$ to the class $C$, as:
    \begin{itemize}
        \item nPMI value of zero indicates no association between token $t$ and class $C$.
        \item A positive nPMI indicates that samples from class $C$ are more likely to use the token $t$.
        \item A negative nPMI indicates that samples from class $C$ are less likely to use the token $t$.
    \end{itemize}
    Furthermore, since we need to find both relevant and irrelevant tokens to the target class $C$, we divide the dataset into samples that have been annotated as \textit{relevant} to the target class $C$ and those with \textit{irrelevant} label. Therefore, we generate two lists of \textit{extracted relevant tokens} and \textit{extracted irrelevant tokens} to the class $C$ by calculating nPMI measure for each token in these two categories. In this way, higher $nPMI(t,C)$ on the relevant samples of class $C$ implies that the token $t$ is more relevant to class $C$, and higher $nPMI(x,C)$ on the irrelevant samples of class $C$ indicates that the token $x$ is less relevant to class $C$.
\end{enumerate}
In the testing phase, we highlight two tokens that are likely more relevant and two tokens that are likely less relevant to the class $C$. For selecting the top two relevant tokens, we use a list of \textit{expert-provided relevant tokens}  in addition to the list \textit{extracted relevant tokens} with the following procedure:
\begin{enumerate}
    \item If the candidate tokens in the test sample exist in the \textit{expert-provided relevant tokens}, they are selected and assigned a relevancy measure $nPMI=1.0$.
    \item We select top-k tokens from \textit{extracted relevant tokens} with the highest nPMI measures.
    \item We merge these two lists and select the top two tokens with the highest nPMI measures from the merged list.
\end{enumerate}
For selecting the top two less-relevant tokens, we use the list \textit{extracted relevant tokens} and select top-k tokens with the highest nPMI measures. Since it may happen that the model has interpreted the same tokens as less-relevant and as more-relevant, to avoid confusion we remove such tokens from the less-relevant list and select the top two tokens.

\subsection{\textbf{Reasoning}}

The Reasoning interface presents explanations for why a single tweet message could be classified to a particular class (\textbf{Why}), as well as reasons for why it might not belong to that class (\textbf{Why Not}). 
Annotators can review these rationales to inform their decision-making process. 
They can evaluate the tweet from both perspectives and then select the class label for that tweet. 
An example of this interface is shown in Figure \ref{fig:fig_prototypes} (b). 

To generate the reasoning, we  
explore  
the capabilities of 
LLMs 
by prompting an LLM model with the message and proper instructions. 
Since LLMs have been trained on a massive corpus they have good knowledge of subjects like \textit{Transportation Means}, but for the annotating task, we need to provide a definition of the subject that represents the annotation task's requirements, such as institutional insights. Moreover, LLMs usually generate long reasonings which can overwhelm the users with details that can confuse the users, so we need to provide short and informative reasoning. The procedure for generating reasoning by LLM is as follows:
\begin{enumerate}
    \item We prompt the LLM with the given tweet and expert-provided definition of class $C$, and ask the LLM to generate the reasoning about why the tweet is relevant or irrelevant to the class $C$.
    \item We extract all sentences from generated reasoning.
    \item Again, we prompt the LLM by providing these extracted sentences as the options and ask the LLM to select the reasoning from these options.
\end{enumerate}


\subsection{\textbf{Context}}

The Context interface leverages insights from disagreements between Experts and Beginners, assisting Beginners in identifying potential errors and uncovering hidden contextual information within the displayed tweet. 
This interface incorporates two key elements: first, it provides cues in the form of hints for keywords that have confused annotators in the past and led to incorrect decisions. 
Second, it reveals the accurate context of the presented tweet, which might not be explicitly stated, by utilizing ground truth data with the assistance of LLM-based summarization techniques, 
as shown in Figure \ref{fig:fig_prototypes} (c). 


For the first hint, we specified the tokens in the given sample that may cause ambiguity for the user to select the correct label. 
In the second element, we utilize feedback and reasoning provided by both Experts and Beginners, gathered during our formative study, with a specific focus on areas where they disagreed. 
To extract ambiguous tokens for a given label, we figure out the tokens from our dataset which are distributed almost equally in both relevant and irrelevant classes. 
However, these tokens need to occur at least \textit{min\_freq} (3 in our experiments) times in each class. 
For measuring the ambiguity of a token $t_k$ in a class $c_i \in \{\textit{relevant}, \textit{irrelevant}\}$, we employ the ambiguity measure $AM(t_k, c_i)$ as described in \cite{mengle2009ambiguity}. $AM(t_k,c_i) = \mathit{tf} (t_k,c_i)/\mathit{tf} (t_k)$, where $\mathit{tf}(t_k, c_i)$ is the frequency of token $t_k$ in class $c_i$, and  $\mathit{tf}(t_k)$ is the frequency of token $t_k$ in all classes. This measure represents the frequency of a token $t_k$ in each class, so to measure the ambiguity of token $t_k$ in the given label, we calculate the ambiguity measure $AM(t_k) = \mathit{max}(AM(t_k, \mathit{relevant}), (AM(t_k, \mathit{irrelevant}))$, which is the maximum of the ambiguity measure for a token in both relevant and irrelevant classes.
If a token $t_k$ equally occurs in both relevant and irrelevant classes, the ambiguity measure $AM(t_k)=0.5$, which means the token has the highest ambiguity. If the token occurs only in one class, the ambiguity measure is equal to 1.0 which implies that the token is indicative of a particular class, so it is not ambiguous at all. Therefore, the range of ambiguity measure is [0.5, 1.0]. We set a threshold \textit{max\_amb} (0.7 in our experiments), and tokens with $AM(t_k) \leq max\_amb$ are considered as ambiguous tokens.
In our experiment, we use the samples from the disagreement part of the dataset, as the training set, to figure out ambiguous tokens for each label (TM or DI), since these samples can represent the knowledge gap between Experts and Beginners. First, we calculate the ambiguity measure for all tokens in the training set and then select the top-k (k=3 in our experiment) token with the lowest $AM$ measure (higher ambiguity).
To provide reasoning about the knowledge gap between users based on their feedback, we prompt an LLM with the test sample (see appendix), selected labels by the users for that sample, and their feedback regarding their choice and ask the model to generate reasoning. We generate the reasoning for a given sample through the following steps:
\begin{enumerate}
    \item For each annotator, we prompt the LLM with the text, the definition of the label, and the annotator's feedback and ask the LLM to generate reasoning behind the annotator's prediction. Providing the definition of the label helps the LLM to consider institutional insights in generating the reasoning.
    \item  We prompt the LLM with the text, the definition of the label, and two reasoning generated in step (1) and ask the model to generate the reason behind the users' disagreement. We change the order of providing annotators' reasoning to eliminate any bias that may caused by the order. 

\end{enumerate}



















%% file: 05_S2.tex
\section{Summative study (S2)}

In this study, we aim to understand how different interfaces can make a difference in reducing the knowledge gap and increasing annotation performance for the less experienced annotators in the disaster management context. 
S2's Research Questions (RQs) are as follows:
\begin{itemize}
    \item \textbf{RQ1.} How do different annotation interface designs impact annotators' behavioral performance in terms of accuracy and efficiency?
    \item \textbf{RQ2.} Which design is perceived as the most effective in addressing the knowledge gap among annotators?
\end{itemize}

\subsection{Method}

\subsubsection{Recruitment}

For recruiting participants, we followed the same approach as our formative study.  
One of the authors with expertise in the field contacted the CERT volunteers both in person and via email, inquiring if they were interested in taking part in the study. 
Those who expressed interest were then selected for the annotation task, resulting in a total of 13 participants. 
For this round, we specifically enlisted Beginners, and the participants are entirely distinct from those involved in the previous study. 
On average the participants have 3.5 years of experience in the field. 
Each participant was compensated with a gift card.

\subsubsection{Data sampling and environment setup}
\label{datasample}

In the final phase of our study, we meticulously selected 459 data points from the earlier disagreement analysis (section 3.4.3). These data points, indicators of disagreements between the experts and novices in disaster-related cases, serve as the basis for our investigation. Our primary aim is to validate whether our proposed designs can narrow the knowledge gap and foster consensus among annotators. To ensure impartiality, we enlisted entirely new participants who had no prior exposure to the dataset.

The study is structured to assess three distinct design conditions, employing a Latin square design for counterbalancing ~\cite{bradley1958complete}. 
To mitigate the influence of a potential learning effect in a within-subject design, we curated three separate datasets, namely D1, D2, and D3, using a stratified sampling approach. Within each interface, we included 40 tweet samples, evenly distributed between Transportation Means (TM) and Damaged Infrastructure (DI) classes. In both classes, 10 samples represented False Positives (FP), while the other 10 represented False Negatives (FN), based on the previous annotation task outcomes. In sum, each participant undertook the annotation of 120 tweets, including three distinct design variations, each comprising 40 tweets.

\subsubsection{Training and annotation exercise}

Before commencing the actual annotation task, a training session was conducted to acquaint our participants with the task's objectives and the various system interfaces. We began by introducing the three interface designs and providing operational instructions. Interactive training ensued, featuring example tweets from their respective class or label. Active participation and decision-sharing were encouraged during this phase. Participants were also trained on system features, such as task tracking, interface navigation, and survey completion.

Participants received individual access to the application through a unique URL and user credentials for the primary annotation task. Approximately 3 hours were allocated for task completion. Each user accessed one dataset at a time, transitioning to a different interface upon completion. The sequence of dataset presentation was individualized and determined via the Latin square technique~\cite{bradley1958complete}.

After annotating each interface, participants completed a mandatory survey assessing efficiency, effectiveness, and knowledge gap reduction, using a Likert scale from 1 (lowest satisfaction) to 7 (highest satisfaction).

\subsection{Results}

Since the summative study relied on ground truth-driven data and was conducted within subjects, achieving consensus among annotators wasn't our primary focus.
\TwoC{One of the common performance measures in HCI research is focusing on dependent variables to understand the impact of design~\cite{10.5555/1841406}.
In experimental research two such variables are, Efficiency- how fast a user can finish a task, and Accuracy- how error-free or precise users are in completing a task~\cite{10.5555/1841406}.}
We established five metrics to gauge the performance of the annotation task and determine which design was most effective in enhancing annotation accuracy and efficiency. 
The initial two metrics assess users' behavioral accuracy and efficiency by examining the outcomes of the annotation exercise. 
The remaining three metrics focus on user perceptions, measuring attitudes regarding accuracy, efficiency, and knowledge gap reduction, as determined through survey responses.

\subsubsection{\textbf{Behavioral Accuracy}}

\begin{figure*}
  \centering
  \includegraphics[width=\linewidth]{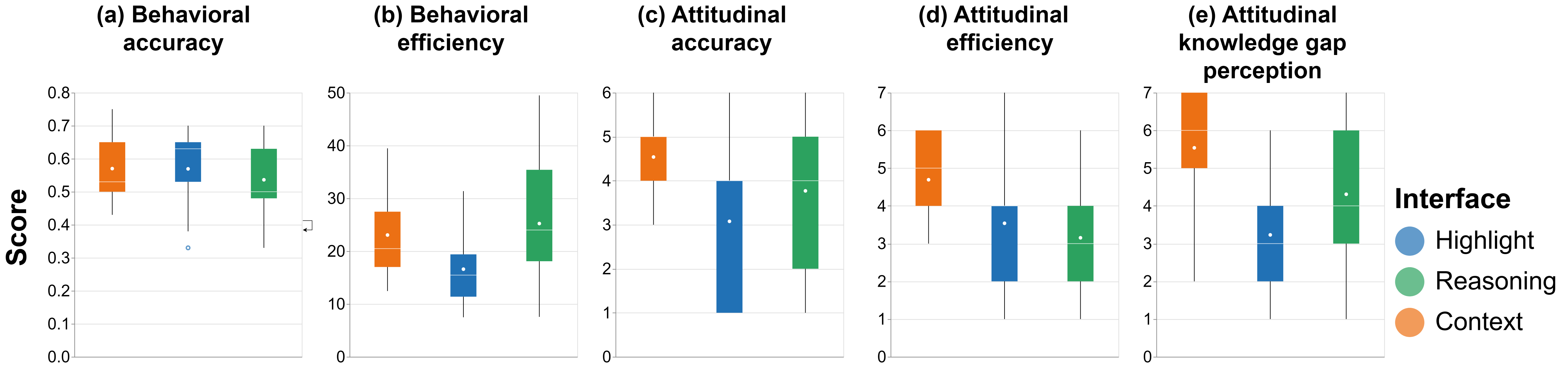}
  \caption{Summative study result under S1 situation - (a) Behavioral Accuracy: box plots for accuracy scores of all users for three interfaces, (b) Behavioral Efficiency: box plots for total completion time (in minutes) of all users for three interfaces, (c) Attitudinal Accuracy: box plots for users perceptual accuracy ratings on a scale of 1 to 7, (d) Attitudinal Efficiency: box plots for users perceptual efficiency ratings on a scale of 1 to 7, (e) Attitudinal Knowledge Gap Perception: box plots for users knowledge gap perception ratings on a scale of 1 to 7}
  \vspace{-0.1in}
  \label{fig:fig_result_sum}
  \Description{final}
\end{figure*}

Behavioral accuracy per user evaluates the correct annotation of samples against the ground truth for a specific interface. To compare accuracy across the three designs, we used the Kruskal-Wallis test and post hoc Dunn analysis. In scenario S1 with all samples considered, the mean accuracy for the Highlight design was 0.57, the Reasoning interface 0.54, and the Context design 0.57, with no significant differences. Figure \ref{fig:fig_result_sum} (a)'s box plot illustrates this, where the white circle indicates the mean. In S2, we excluded samples with less than 5 seconds spent per question due to accidental clicks without reading the tweet or cues. 
For example, a participant mentioned, \textit{``Answers require clicking on very small (at least on my screen) radio buttons. I could not select an option by clicking on the text next to the radio buttons or vicinity. This led me to make a mistake on one of the questions''}. 
Another point of consideration is that the average time taken per question by all users exceeds 5 seconds, supporting the exclusion of those samples. In situation S2, the mean accuracy values shifted for the Highlight interface to 0.56 and for the Context interface to 0.58, while no change was noted for the Reasoning interface. 
We examined another scenario, S3, in which we excluded one participant's data from the evaluation due to their significantly shorter time spent compared to all other annotators. 
For S3, the mean accuracy scores updated to 0.55 for the Highlight interface, 0.52 for the Reasoning interface, and 0.58 for the Context interface. 
However, in all these scenarios, no statistically significant differences were observed among the three design conditions and in the pair-wise comparisons. 

From the box plot, it is clear that the median value for the Highlight interface surpasses the mean, suggesting that some participants achieved exceptionally high performance with this interface. 
However, there is a notable difference between the maximum and minimum range of performance outcomes for Highlight and Reasoning interfaces, which indicates some participants achieved noticeably lower performance using these interfaces as well. 
Both min and max accuracy scores for the Context interface are 
better than those of the other two designs.

We further explored using the Chi-square test (contingency table) and conducted a question-wise assessment for all participants across the three designs in scenario S1. 
Since each interface comprised 40 tweets, we designated the initial tweet in all interfaces as question 1, the second as question 2, and so forth. 
We count the accuracy scores for each question for all participants. 
In our findings, the Context interface outperformed both the Highlight and Reasoning interfaces for 13 questions and outperformed one interface for 3 questions. 
Among these questions, 7 cases exhibited significant differences compared to the Highlight and Reasoning interfaces (p < 0.04). 
Similarly, the Highlight interface outperformed two designs for 11 questions and outperformed one interface for one question. 
6 cases among them showed significant differences (p < 0.03). 
Lastly, the Reasoning interface demonstrated better accuracy in 10 questions compared to the other two designs and outperformed one design for 4 questions. 
We observed 3 cases being significantly better (p < 0.03). 
For two questions all three designs exhibited similar accuracy scores having no impact on each other. 

\subsubsection{\textbf{Behavioral Efficiency}}

Behavioral efficiency measures the speed and completion time of annotators using different interface types. 
The aim is to measure which interface is efficient in providing information and allowing annotators to quickly annotate or label data with minimal time and effort.  
When considering data from all annotators (S1), the mean completion time for annotating 40 tweets was 16.59 minutes for the Highlight design, 25.23 minutes for the Reasoning interface, and 23.05 minutes for the Context design. 
Notably, the Highlight interface appeared to be the fastest in terms of completion time. 
Although there was no statistical significance observed when comparing the speed across all three design conditions, we detected significant differences in pair-wise comparisons. 
Specifically, the Highlight interface proved to be significantly faster than both the Reasoning interface (p < 0.05) and the Context interface (p < 0.03). 

In scenario S2, where we excluded samples with less than 5 seconds spent per tweet, the mean completion times remained relatively stable: 25.21 minutes for the Highlight design, 25.21 minutes for Reasoning, and no change for the Context design. 
 In scenario S3, the average completion times were increased across all interfaces. Specifically, for the Highlight design, the mean completion time extended to 17.35 minutes, while for Reasoning, it reached 26.70 minutes, and for the Context interface, it was 23.93 minutes.
 The Reasoning interface captures the majority of users' time. 
 In both S2, S3 the Highlight interface stood out as significantly faster (p < 0.03) than the other two designs. 
This occurred mainly because annotators didn't need to invest additional time in reading AI-generated explanations, allowing them to complete the task more swiftly.

\subsubsection{\textbf{Attitudinal Accuracy}}

Attitudinal accuracy assesses the perceived accuracy of annotators based on their survey responses. 
We posed a specific question for this metric: ``I found the way the current interface provides information enables accurate annotation decisions with less error'', aiming to assess the annotators' subjective perception of the interface's ability to support them in making correct and error-free annotation decisions. 
Respondents used a Likert scale ranging from 1 (strongly disagree) to 7 (strongly agree) to rate their agreement. 
The mean attitudinal accuracy score for the Highlight interface was 3.08, for the Reasoning interface it was 3.77, and for the Context interface, it reached 4.54 out of 7. 
It's noteworthy that annotators believed the Context design facilitated more accurate decision-making compared to the other two interfaces, even though there appeared to be no significant behavioral accuracy differences between the Highlight and Context designs. 
The agreement scores for the Highlight feature exhibit significant variability, with some participants assigning a score of 1 while others rated it as 7. 
A similar pattern of variability is observed for the Reasoning interface. In contrast, for the Context design, the range of agreement scores falls between 3 and 6. 
From the Kruskal-Willis analysis, we found a significant statistical difference between these three conditions where Context design achieved a better attitudinal accuracy score than other designs. 

\subsubsection{\textbf{Attitudinal Efficiency}}

To observe our annotators' perceptual efficiency, we asked a question in the survey, ``I found the way the current interface provides information, enables fast annotation using less time''. 
The goal is to understand whether the participants found the interface to be time-saving and efficient in its information presentation and annotation process. 
The mean attitudinal efficiency scores were as follows: 3.54 for the Highlight interface, 3.15 for the Reasoning interface, and 4.69 for the Context interface, on a scale of 1 to 7. 
The Reasoning interface achieved the lowest score, aligning with our observations from behavioral efficiency, where annotators took the longest time to complete tasks using this interface due to the additional time spent reading AI-generated explanations. 
Surprisingly, while the Highlight design was found to be the fastest, annotators tended to perceive the Context design as the most efficient for facilitating effective decision-making. 
This demonstrates that users have a strong preference for the Context interface (p < 0.04).

\subsubsection{\textbf{Attitudinal Knowledge Gap Perception}}

Our last survey question was ``I found the way the current interface provides information, helps me learn the aspect that I could overlook otherwise'', intends to determine whether the way the interface presents information assists annotators in learning and understanding certain aspects they might have missed or overlooked otherwise while decision-making. 
The mean survey scores were as follows: 3.23 for the Highlight interface, 4.31 for the Reasoning interface, and 5.54 for the Context interface, on a scale of 1 to 7. 
Statistical significant differences were observed among these conditions (p < 0.03), suggesting that annotators perceived the Context design as offering more valuable support compared to the other designs. 
As one of the annotators mentioned,\textit{ ``I found both, the context and reasoning options more useful than the highlighting. On one occasion I answered a question without looking at the helping text, then I read the helping text and changed my response for better, I hope. However, the design where I had both, the keywords on the tweet and helping text options, may have helped me speed up my responses''}. 
Interestingly, the Highlight design received the lowest score in terms of attitudinal metrics, despite its behavioral metrics showing the opposite trend. 

%% file: 06_implications_for_design.tex
\section{Discussion \& Implications for Design}

This section provides insights we learned from S1 and S2 that can motivate future research and annotation user interface designs.
We will discuss how the annotation user interface design can be improved by leveraging the two notions of confusing words and hidden context. In discussing the possible expansion, we will first provide how the two can be applied depending on the data type. 
Next, we will discuss how the two can be applied to different domains in disaster management. 
Finally, we discuss how an advanced technical pipeline can be considered to advance design for closing the knowledge gap using confusing words and hidden context.

\subsection{Design directions}
Advanced annotation designs can be adopted based on the data type and modality, focusing on addressing ambiguous elements and hidden contexts. 
Several potential directions in this regard are outlined below.

\subsubsection{Adopting in multi-step annotation workflows}
Unlike microtasks, advanced task requires comprehensive reasoning based on domain knowledge. 
In that sense, the two notions can be applied to the interface to help annotators apply the ``divide-and-conquer'' approach in annotation.
This line of designs might consider implementing a step-wise~\cite{Stureborg_2023}, where the workflows will guide annotators through a sequence of stages. 
For instance, In the first step, annotators can focus on identifying the hidden context and highlighting ambiguous elements. 
In subsequent steps, they can provide annotations based on the clarified context, leading to more accurate and comprehensive annotations. 
While such designs can help annotators to be comprehensive in checking multifaceted aspects of the annotation, the thread of this approach is possible expansion of spending time.
Annotations might be guided through additional explanations, links to external resources, or suggestions for seeking additional information. 
After the detailed clarification step, there can be a resolution step where the system offers guidance by suggesting strategies for disambiguation, such as considering the surrounding context or consulting domain-specific knowledge. 

\subsubsection{Applying the notions to images and videos}
Our study findings show that confusing words and hidden context can benefit annotation research in images and videos. 
Computer vision models are increasingly becoming contextualized and applied in professional tasks. 
Confusing ``patterns'', for example, can be applied in detecting falsely correlated objects in classification or object detection models--where the object type has a strong correlation to a particular class or object type but itself doesn't mean that the data points should be classified to that object (e.g., tennis racquets or baseball bats in gender classifier~\cite{gao2022aligning}).
Hidden context can also help provide underrepresented knowledge in varying domains, such as damage assessment or medical imaging areas. 

\subsubsection{Adding feedback loops for improving context and ambiguous word lists}
Introducing feedback loops from either domain experts or trained AI models into the annotation process can offer annotators personalized feedback tailored to their annotation performance~\cite{gao2022res, sun2023designing}. 
These suggestions may originate from domain experts who can provide insights into beginners' decisions or be delivered through AI agents. 
Such feedback can pinpoint their strengths and areas requiring improvement including their ability to handle hidden contexts and ambiguities. 
This feedback can take the form of comments, ratings, or suggested improvements directly within the annotation interface. 
The system collects and aggregates feedback from multiple annotators or analyzes them to identify common themes. 
Feedback integration can be done in real-time or iteration basis,  annotators can contribute to an evolving set of hints.



\subsubsection{Multimodal contextualization and ambiguity detection}

In a multimodal interface, the concept of explaining hidden context can be extended to include not only textual but also visual or auditory context. 
For example, if annotating an image with text, the interface could provide explanations for why certain text elements were chosen or why specific regions of the image are relevant. 
For instance, one of our participants mentioned, \textit{``There were links that we didn't get to see or open in the tweet. So if I could have seen the link then that might have given me more context''}. 
The system should be capable of identifying and highlighting ambiguous elements not only in text but also in visual or auditory forms. 
Offering alternative word choices, explaining ambiguous visual elements, transcribing the speech, or clarifying confusing words could be useful to the context. 
This might involve using a combination of NLP techniques for textual content and computer vision or audio processing techniques for other modalities.

\subsection{Application in disaster management}
The scope, magnitude, and complexity of disasters influence social media use and value~\cite{hughespeterson2014}. On the ground decision-making is driven by these disaster characteristics and 
 the relevance of 
 social media content to support such decision-making processes for response across different sectors 
 (e.g., transportation) rapidly changes. 
Thus, annotation tasks on such social media content for disaster data analytics systems could face 
varying relevance of words and context in messages for response sectors and lose value as operations transition from response to recovery. 
As a consequence of these dynamics, combined with early indications in our S1 study that annotation tasks may have been complex and overwhelming for beginning annotators, the future design could focus on annotation aids for microtasks with binary classifications. 

\subsection{Advancing technical pipeline}

Our experimental interfaces highlighted both technical challenges and opportunities for future designs. 
The dataset's size, used for extracting relevant and ambiguous words in the Highlight and Context interfaces, can impact accuracy. Expanding the dataset size could enhance method performance. Additionally, LLMs 
have shown 
powerful capabilities in 
NLP tasks such as keyword extraction and generation. Thus, utilizing LLMs with 
appropriate prompts, which capture 
the context of the task, can be an alternative approach 
to generate a list of relevant and ambiguous words for a target label (e.g., Transportation means) based on a given message. Future designs should consider the potential bias in the generation process of LLMs. 
\Four{There is a concern that explanations generated by LLM may be overly verbose or detailed. 
This verbosity can potentially overwhelm users with information.
If explanations are too extensive, users might find it challenging to digest the information quickly leading to cognitive overload and may hinder the decision-making process rather than aiding it. 
Utilizing prompt templates that enhance LLM faithfulness to account for contextual knowledge can improve reasoning quality in future tasks~\cite{zhou-etal-2023-context}.
This information helps the LLM in understanding the context and criteria for different labels, enabling it to make more informed decisions during reasoning tasks. 
However, while including label definitions aids the LLM in reasoning, it does not guarantee that the model will consistently adhere to these definitions. 
Deviation from the provided definitions may occur due to various factors, such as biases in the training data or the model's inherent tendencies.} 
Furthermore, the lessons of ambiguous words and hidden context in the proposed interface designs are focused on single modality of data, i.e., text could inspire the technical implementation of the annotation aids for multimodal annotation tasks. %
The multimodality of messages including images and videos would require the detection of relevant and ambiguous objects to support annotation aids on the interfaces. 
The advanced models using transformers~\cite{khan2022transformers} have shown remarkable capabilities for computer vision tasks that could be leveraged for such annotation aids to support interfaces that aim to reduce knowledge gap on multimodal annotation tasks.

%% file: 07_conclusion.tex
\section{Limitation \& Conclusion}

\Four{Regarding the limitations of this work,} we utilized the same dataset for both testing the interface and generating context to address disagreements between Experts and Beginners in the context interface.
While this may not align with real-world scenarios, we adopted this approach to create a suitable context for our experimental condition. 
In practical applications, an accurate model trained on historical annotated data can be employed to generate a context explanation for resolving user disagreements.
\FourX{Additionally, the S1 focused only on one event type due to time and cost limitations, employing a single-step expert verification for ground truth preparation.
Classifying our participants as experts and beginners could produce unaccounted variability among the Beginner annotators which could affect generalizability. 
Incorporating multivariate data and preparing ground truth with additional feedback from multiple domain experts could broaden applicability.}

Our study findings underscore the critical nature of annotation tasks in disaster management, which can introduce ambiguity for annotators across varying levels of expertise. 
We investigate the potential of AI-assisted interface designs to mitigate the knowledge gap among less experienced annotators. The empirical study reveals that the most prevalent reasons for disagreements are confusing words and hidden context for textual annotation tasks. 
In response, a novel interface is introduced that provides cues to address these challenges. 
In our future work, we plan to expand this research to accommodate multimodal data annotation tasks and enhance the implemented technical pipeline's 
predictive capabilities to help generate aids for further improving the annotation performance.

%% file: appendix.tex
\section{Appendix A: Additional Figures}

\begin{figure}[!htb]
  \centering
\fbox{\begin{minipage}{\linewidth}

\textbf{Instruction:} Read the definition and the annotator's reasoning about the following tweet and complete the answer. 

\textbf{Definition:} <given-definition>

\textbf{Tweet:} <given-tweet>

\textbf{Annotator:} <annotator-feedback>

\textbf{Answer:} Annotator believes that the tweet is <annotator-prediction> to <disagreement-label> because

\end{minipage}}
\caption{Prompt for generating reasoning based on the user's feedback}
\label{fig:prompt-reasoning-generator}
\end{figure}

\begin{figure}[!htb]
  \centering
\fbox{\begin{minipage}{\linewidth}

\textbf{Instruction:} Two annotators generated the following reasoning for the following task and tweet. What is the reason for their disagreement? 

\textbf{Task:} Based on the following definition, is the following tweet relevant to <disagreement-label>?

\textbf{Definition:} <given-definition>

\textbf{Tweet}: <given-tweet>

\textbf{Annotator 1:} <annotator1-reasoning>

\textbf{Annotator 2:} <annotator2-reasoning>

\textbf{Answer:} The reason for their disagreement is that

\end{minipage}}
\caption{Prompts for generating reasoning behind the users' disagreement}
\label{fig:prompt-context-generator}
\end{figure}




%% file: 00_anchor.bbl

\begin{thebibliography}{63}


\ifx \showCODEN    \undefined \def \showCODEN     #1{\unskip}     \fi
\ifx \showDOI      \undefined \def \showDOI       #1{#1}\fi
\ifx \showISBNx    \undefined \def \showISBNx     #1{\unskip}     \fi
\ifx \showISBNxiii \undefined \def \showISBNxiii  #1{\unskip}     \fi
\ifx \showISSN     \undefined \def \showISSN      #1{\unskip}     \fi
\ifx \showLCCN     \undefined \def \showLCCN      #1{\unskip}     \fi
\ifx \shownote     \undefined \def \shownote      #1{#1}          \fi
\ifx \showarticletitle \undefined \def \showarticletitle #1{#1}   \fi
\ifx \showURL      \undefined \def \showURL       {\relax}        \fi
\providecommand\bibfield[2]{#2}
\providecommand\bibinfo[2]{#2}
\providecommand\natexlab[1]{#1}
\providecommand\showeprint[2][]{arXiv:#2}

\bibitem[Ackerman(2000)]%
        {ackerman2000intellectual}
\bibfield{author}{\bibinfo{person}{Mark~S Ackerman}.} \bibinfo{year}{2000}\natexlab{}.
\newblock \showarticletitle{The intellectual challenge of CSCW: the gap between social requirements and technical feasibility}.
\newblock \bibinfo{journal}{\emph{Human--Computer Interaction}} \bibinfo{volume}{15}, \bibinfo{number}{2-3} (\bibinfo{year}{2000}), \bibinfo{pages}{179--203}.
\newblock


\bibitem[Amershi and Morris(2008)]%
        {amershi2008cosearch}
\bibfield{author}{\bibinfo{person}{Saleema Amershi} {and} \bibinfo{person}{Meredith~Ringel Morris}.} \bibinfo{year}{2008}\natexlab{}.
\newblock \showarticletitle{CoSearch: a system for co-located collaborative web search}. In \bibinfo{booktitle}{\emph{Proceedings of the SIGCHI conference on human factors in computing systems}}. \bibinfo{pages}{1647--1656}.
\newblock


\bibitem[Ara and Hashemi(2021a)]%
        {ara2021ride}
\bibfield{author}{\bibinfo{person}{Zinat Ara} {and} \bibinfo{person}{Mahdi Hashemi}.} \bibinfo{year}{2021}\natexlab{a}.
\newblock \showarticletitle{Ride hailing service demand forecast by integrating convolutional and recurrent neural networks}. In \bibinfo{booktitle}{\emph{Proceedings of the 33rd International Conference on Software Engineering and Knowledge Engineering}}. \bibinfo{pages}{463--468}.
\newblock


\bibitem[Ara and Hashemi(2021b)]%
        {ara2021traffic}
\bibfield{author}{\bibinfo{person}{Zinat Ara} {and} \bibinfo{person}{Mahdi Hashemi}.} \bibinfo{year}{2021}\natexlab{b}.
\newblock \showarticletitle{Traffic Flow Prediction using Long Short-Term Memory Network and Optimized Spatial Temporal Dependencies}. In \bibinfo{booktitle}{\emph{2021 IEEE International Conference on Big Data (Big Data)}}. IEEE, \bibinfo{pages}{1550--1557}.
\newblock


\bibitem[Ashktorab et~al\mbox{.}(2021)]%
        {ashktorab2021ai}
\bibfield{author}{\bibinfo{person}{Zahra Ashktorab}, \bibinfo{person}{Michael Desmond}, \bibinfo{person}{Josh Andres}, \bibinfo{person}{Michael Muller}, \bibinfo{person}{Narendra~Nath Joshi}, \bibinfo{person}{Michelle Brachman}, \bibinfo{person}{Aabhas Sharma}, \bibinfo{person}{Kristina Brimijoin}, \bibinfo{person}{Qian Pan}, \bibinfo{person}{Christine~T Wolf}, {et~al\mbox{.}}} \bibinfo{year}{2021}\natexlab{}.
\newblock \showarticletitle{Ai-assisted human labeling: Batching for efficiency without overreliance}.
\newblock \bibinfo{journal}{\emph{Proceedings of the ACM on Human-Computer Interaction}} \bibinfo{volume}{5}, \bibinfo{number}{CSCW1} (\bibinfo{year}{2021}), \bibinfo{pages}{1--27}.
\newblock


\bibitem[Austin(2003)]%
        {austin2003transactive}
\bibfield{author}{\bibinfo{person}{John~R Austin}.} \bibinfo{year}{2003}\natexlab{}.
\newblock \showarticletitle{Transactive memory in organizational groups: the effects of content, consensus, specialization, and accuracy on group performance.}
\newblock \bibinfo{journal}{\emph{Journal of applied psychology}} \bibinfo{volume}{88}, \bibinfo{number}{5} (\bibinfo{year}{2003}), \bibinfo{pages}{866}.
\newblock


\bibitem[Bai et~al\mbox{.}(2022)]%
        {bai2022training}
\bibfield{author}{\bibinfo{person}{Yuntao Bai}, \bibinfo{person}{Andy Jones}, \bibinfo{person}{Kamal Ndousse}, \bibinfo{person}{Amanda Askell}, \bibinfo{person}{Anna Chen}, \bibinfo{person}{Nova DasSarma}, \bibinfo{person}{Dawn Drain}, \bibinfo{person}{Stanislav Fort}, \bibinfo{person}{Deep Ganguli}, \bibinfo{person}{Tom Henighan}, \bibinfo{person}{Nicholas Joseph}, \bibinfo{person}{Saurav Kadavath}, \bibinfo{person}{Jackson Kernion}, \bibinfo{person}{Tom Conerly}, \bibinfo{person}{Sheer El-Showk}, \bibinfo{person}{Nelson Elhage}, \bibinfo{person}{Zac Hatfield-Dodds}, \bibinfo{person}{Danny Hernandez}, \bibinfo{person}{Tristan Hume}, \bibinfo{person}{Scott Johnston}, \bibinfo{person}{Shauna Kravec}, \bibinfo{person}{Liane Lovitt}, \bibinfo{person}{Neel Nanda}, \bibinfo{person}{Catherine Olsson}, \bibinfo{person}{Dario Amodei}, \bibinfo{person}{Tom Brown}, \bibinfo{person}{Jack Clark}, \bibinfo{person}{Sam McCandlish}, \bibinfo{person}{Chris Olah}, \bibinfo{person}{Ben Mann}, {and} \bibinfo{person}{Jared
  Kaplan}.} \bibinfo{year}{2022}\natexlab{}.
\newblock \bibinfo{title}{Training a Helpful and Harmless Assistant with Reinforcement Learning from Human Feedback}.
\newblock
\newblock
\showeprint[arxiv]{2204.05862}~[cs.CL]


\bibitem[Bartolo et~al\mbox{.}(2020)]%
        {bartolo2020beat}
\bibfield{author}{\bibinfo{person}{Max Bartolo}, \bibinfo{person}{Alastair Roberts}, \bibinfo{person}{Johannes Welbl}, \bibinfo{person}{Sebastian Riedel}, {and} \bibinfo{person}{Pontus Stenetorp}.} \bibinfo{year}{2020}\natexlab{}.
\newblock \showarticletitle{{Beat the AI: Investigating adversarial human annotation for reading comprehension}}.
\newblock \bibinfo{journal}{\emph{Transactions of the Association for Computational Linguistics}}  \bibinfo{volume}{8} (\bibinfo{year}{2020}), \bibinfo{pages}{662--678}.
\newblock


\bibitem[Bentley and Peesapati(2017)]%
        {bentley2017searchmessenger}
\bibfield{author}{\bibinfo{person}{Frank~R Bentley} {and} \bibinfo{person}{S~Tejaswi Peesapati}.} \bibinfo{year}{2017}\natexlab{}.
\newblock \showarticletitle{SearchMessenger: Exploring the use of search and card sharing in a messaging application}. In \bibinfo{booktitle}{\emph{Proceedings of the 2017 ACM Conference on Computer Supported Cooperative Work and Social Computing}}. \bibinfo{pages}{1946--1956}.
\newblock


\bibitem[Bouma(2009)]%
        {bouma2009normalized}
\bibfield{author}{\bibinfo{person}{Gerlof Bouma}.} \bibinfo{year}{2009}\natexlab{}.
\newblock \showarticletitle{Normalized (pointwise) mutual information in collocation extraction}.
\newblock \bibinfo{journal}{\emph{Proceedings of GSCL}}  \bibinfo{volume}{30} (\bibinfo{year}{2009}), \bibinfo{pages}{31--40}.
\newblock


\bibitem[Brachman et~al\mbox{.}(2022)]%
        {10.1145/3555212}
\bibfield{author}{\bibinfo{person}{Michelle Brachman}, \bibinfo{person}{Zahra Ashktorab}, \bibinfo{person}{Michael Desmond}, \bibinfo{person}{Evelyn Duesterwald}, \bibinfo{person}{Casey Dugan}, \bibinfo{person}{Narendra~Nath Joshi}, \bibinfo{person}{Qian Pan}, {and} \bibinfo{person}{Aabhas Sharma}.} \bibinfo{year}{2022}\natexlab{}.
\newblock \showarticletitle{Reliance and Automation for Human-AI Collaborative Data Labeling Conflict Resolution}.
\newblock \bibinfo{journal}{\emph{Proc. ACM Hum.-Comput. Interact.}} \bibinfo{volume}{6}, \bibinfo{number}{CSCW2}, Article \bibinfo{articleno}{321} (\bibinfo{date}{nov} \bibinfo{year}{2022}), \bibinfo{numpages}{27}~pages.
\newblock
\urldef\tempurl%
\url{https://doi.org/10.1145/3555212}
\showDOI{\tempurl}


\bibitem[Bradley(1958)]%
        {bradley1958complete}
\bibfield{author}{\bibinfo{person}{James~V Bradley}.} \bibinfo{year}{1958}\natexlab{}.
\newblock \showarticletitle{{Complete counterbalancing of immediate sequential effects in a Latin square design}}.
\newblock \bibinfo{journal}{\emph{J. Amer. Statist. Assoc.}} \bibinfo{volume}{53}, \bibinfo{number}{282} (\bibinfo{year}{1958}).
\newblock
\urldef\tempurl%
\url{https://www.tandfonline.com/doi/abs/10.1080/01621459.1958.10501456}
\showURL{%
\tempurl}


\bibitem[Cabrera et~al\mbox{.}(2006)]%
        {cabrera2006determinants}
\bibfield{author}{\bibinfo{person}{Angel Cabrera}, \bibinfo{person}{William~C Collins}, {and} \bibinfo{person}{Jesus~F Salgado}.} \bibinfo{year}{2006}\natexlab{}.
\newblock \showarticletitle{Determinants of individual engagement in knowledge sharing}.
\newblock \bibinfo{journal}{\emph{The International Journal of Human Resource Management}} \bibinfo{volume}{17}, \bibinfo{number}{2} (\bibinfo{year}{2006}), \bibinfo{pages}{245--264}.
\newblock


\bibitem[Cao and Choe(2020)]%
        {cao2020building}
\bibfield{author}{\bibinfo{person}{Quoc~Dung Cao} {and} \bibinfo{person}{Youngjun Choe}.} \bibinfo{year}{2020}\natexlab{}.
\newblock \showarticletitle{{Building damage annotation on post-hurricane satellite imagery based on convolutional neural networks}}.
\newblock \bibinfo{journal}{\emph{Natural Hazards}} \bibinfo{volume}{103}, \bibinfo{number}{3} (\bibinfo{year}{2020}), \bibinfo{pages}{3357--3376}.
\newblock


\bibitem[Carlotto(2009)]%
        {10.1080/01431160802672864}
\bibfield{author}{\bibinfo{person}{Mark~J. Carlotto}.} \bibinfo{year}{2009}\natexlab{}.
\newblock \showarticletitle{Effect of errors in ground truth on classification accuracy}.
\newblock \bibinfo{journal}{\emph{International Journal of Remote Sensing}} \bibinfo{volume}{30}, \bibinfo{number}{18} (\bibinfo{year}{2009}), \bibinfo{pages}{4831--4849}.
\newblock
\urldef\tempurl%
\url{https://doi.org/10.1080/01431160802672864}
\showDOI{\tempurl}
\showeprint{https://doi.org/10.1080/01431160802672864}


\bibitem[Cartwright et~al\mbox{.}(2019)]%
        {10.1145/3290605.3300522}
\bibfield{author}{\bibinfo{person}{Mark Cartwright}, \bibinfo{person}{Graham Dove}, \bibinfo{person}{Ana~Elisa M\'{e}ndez~M\'{e}ndez}, \bibinfo{person}{Juan~P. Bello}, {and} \bibinfo{person}{Oded Nov}.} \bibinfo{year}{2019}\natexlab{}.
\newblock \showarticletitle{Crowdsourcing Multi-Label Audio Annotation Tasks with Citizen Scientists}. In \bibinfo{booktitle}{\emph{Proceedings of the 2019 CHI Conference on Human Factors in Computing Systems}} (Glasgow, Scotland Uk) \emph{(\bibinfo{series}{CHI '19})}. \bibinfo{publisher}{Association for Computing Machinery}, \bibinfo{address}{New York, NY, USA}, \bibinfo{pages}{1–11}.
\newblock
\showISBNx{9781450359702}
\urldef\tempurl%
\url{https://doi.org/10.1145/3290605.3300522}
\showDOI{\tempurl}


\bibitem[Chang et~al\mbox{.}(2017)]%
        {10.1145/3025453.3026044}
\bibfield{author}{\bibinfo{person}{Joseph~Chee Chang}, \bibinfo{person}{Saleema Amershi}, {and} \bibinfo{person}{Ece Kamar}.} \bibinfo{year}{2017}\natexlab{}.
\newblock \showarticletitle{Revolt: Collaborative Crowdsourcing for Labeling Machine Learning Datasets}. In \bibinfo{booktitle}{\emph{Proceedings of the 2017 CHI Conference on Human Factors in Computing Systems}} (Denver, Colorado, USA) \emph{(\bibinfo{series}{CHI '17})}. \bibinfo{publisher}{Association for Computing Machinery}, \bibinfo{address}{New York, NY, USA}, \bibinfo{pages}{2334–2346}.
\newblock
\showISBNx{9781450346559}
\urldef\tempurl%
\url{https://doi.org/10.1145/3025453.3026044}
\showDOI{\tempurl}


\bibitem[Chen et~al\mbox{.}(2021)]%
        {Chen_2021}
\bibfield{author}{\bibinfo{person}{Quan~Ze Chen}, \bibinfo{person}{Daniel~S. Weld}, {and} \bibinfo{person}{Amy~X. Zhang}.} \bibinfo{year}{2021}\natexlab{}.
\newblock \showarticletitle{Goldilocks: Consistent Crowdsourced Scalar Annotations with Relative Uncertainty}.
\newblock \bibinfo{journal}{\emph{Proceedings of the {ACM} on Human-Computer Interaction}} \bibinfo{volume}{5}, \bibinfo{number}{{CSCW}2} (\bibinfo{date}{oct} \bibinfo{year}{2021}), \bibinfo{pages}{1--25}.
\newblock
\urldef\tempurl%
\url{https://doi.org/10.1145/3476076}
\showDOI{\tempurl}


\bibitem[Choi et~al\mbox{.}(2019)]%
        {choi2019aila}
\bibfield{author}{\bibinfo{person}{Minsuk Choi}, \bibinfo{person}{Cheonbok Park}, \bibinfo{person}{Soyoung Yang}, \bibinfo{person}{Yonggyu Kim}, \bibinfo{person}{Jaegul Choo}, {and} \bibinfo{person}{Sungsoo~Ray Hong}.} \bibinfo{year}{2019}\natexlab{}.
\newblock \showarticletitle{{Aila: Attentive interactive labeling assistant for document classification through attention-based deep neural networks}}. In \bibinfo{booktitle}{\emph{Proceedings of the 2019 CHI conference on human factors in computing systems}}. \bibinfo{pages}{1--12}.
\newblock
\urldef\tempurl%
\url{https://doi.org/10.1145/3290605.3300460}
\showURL{%
\tempurl}


\bibitem[Chung et~al\mbox{.}(2019)]%
        {chung2019efficient}
\bibfield{author}{\bibinfo{person}{John Joon~Young Chung}, \bibinfo{person}{Jean~Y Song}, \bibinfo{person}{Sindhu Kutty}, \bibinfo{person}{Sungsoo Hong}, \bibinfo{person}{Juho Kim}, {and} \bibinfo{person}{Walter~S Lasecki}.} \bibinfo{year}{2019}\natexlab{}.
\newblock \showarticletitle{{Efficient elicitation approaches to estimate collective crowd answers}}.
\newblock \bibinfo{journal}{\emph{Proceedings of the ACM on Human-Computer Interaction}} \bibinfo{volume}{3}, \bibinfo{number}{CSCW} (\bibinfo{year}{2019}), \bibinfo{pages}{1--25}.
\newblock


\bibitem[FEMA(2022)]%
        {cert}
\bibfield{author}{\bibinfo{person}{U.S. Department of Homeland~Security FEMA}.} \bibinfo{year}{2022}\natexlab{}.
\newblock \bibinfo{title}{Community Emergency Response Team (CERT)}.
\newblock
\newblock
\urldef\tempurl%
\url{https://www.ready.gov/cert}
\showURL{%
\tempurl}
\newblock
\shownote{April 19}.


\bibitem[Gao et~al\mbox{.}(2021)]%
        {gao2021gnes}
\bibfield{author}{\bibinfo{person}{Yuyang Gao}, \bibinfo{person}{Tong Sun}, \bibinfo{person}{Rishab Bhatt}, \bibinfo{person}{Dazhou Yu}, \bibinfo{person}{Sungsoo Hong}, {and} \bibinfo{person}{Liang Zhao}.} \bibinfo{year}{2021}\natexlab{}.
\newblock \showarticletitle{Gnes: Learning to explain graph neural networks}. In \bibinfo{booktitle}{\emph{2021 IEEE International Conference on Data Mining (ICDM)}}. IEEE, \bibinfo{pages}{131--140}.
\newblock


\bibitem[Gao et~al\mbox{.}(2022a)]%
        {gao2022res}
\bibfield{author}{\bibinfo{person}{Yuyang Gao}, \bibinfo{person}{Tong~Steven Sun}, \bibinfo{person}{Guangji Bai}, \bibinfo{person}{Siyi Gu}, \bibinfo{person}{Sungsoo~Ray Hong}, {and} \bibinfo{person}{Zhao Liang}.} \bibinfo{year}{2022}\natexlab{a}.
\newblock \showarticletitle{Res: A robust framework for guiding visual explanation}. In \bibinfo{booktitle}{\emph{Proceedings of the 28th ACM SIGKDD Conference on Knowledge Discovery and Data Mining}}. \bibinfo{pages}{432--442}.
\newblock


\bibitem[Gao et~al\mbox{.}(2022b)]%
        {gao2022aligning}
\bibfield{author}{\bibinfo{person}{Yuyang Gao}, \bibinfo{person}{Tong~Steven Sun}, \bibinfo{person}{Liang Zhao}, {and} \bibinfo{person}{Sungsoo~Ray Hong}.} \bibinfo{year}{2022}\natexlab{b}.
\newblock \showarticletitle{Aligning eyes between humans and deep neural network through interactive attention alignment}.
\newblock \bibinfo{journal}{\emph{Proceedings of the ACM on Human-Computer Interaction}} \bibinfo{volume}{6}, \bibinfo{number}{CSCW2} (\bibinfo{year}{2022}), \bibinfo{pages}{1--28}.
\newblock


\bibitem[Gooding et~al\mbox{.}(2023)]%
        {gooding2023study}
\bibfield{author}{\bibinfo{person}{Sian Gooding}, \bibinfo{person}{Lucas Werner}, {and} \bibinfo{person}{Victor C{\u{a}}rbune}.} \bibinfo{year}{2023}\natexlab{}.
\newblock \showarticletitle{A Study on Annotation Interfaces for Summary Comparison}. In \bibinfo{booktitle}{\emph{Proceedings of the 17th Linguistic Annotation Workshop (LAW-XVII)}}. \bibinfo{pages}{179--187}.
\newblock


\bibitem[Hong et~al\mbox{.}(2018a)]%
        {hong2018collaborative}
\bibfield{author}{\bibinfo{person}{Sungsoo Hong}, \bibinfo{person}{Minhyang Suh}, \bibinfo{person}{Nathalie Henry~Riche}, \bibinfo{person}{Jooyoung Lee}, \bibinfo{person}{Juho Kim}, {and} \bibinfo{person}{Mark Zachry}.} \bibinfo{year}{2018}\natexlab{a}.
\newblock \showarticletitle{{Collaborative dynamic queries: Supporting distributed small group decision-making}}. In \bibinfo{booktitle}{\emph{Proceedings of the 2018 CHI Conference on Human Factors in Computing Systems}}. \bibinfo{pages}{1--12}.
\newblock
\urldef\tempurl%
\url{https://dl.acm.org/doi/abs/10.1145/3173574.3173640}
\showURL{%
\tempurl}


\bibitem[Hong et~al\mbox{.}(2019b)]%
        {hong2019design}
\bibfield{author}{\bibinfo{person}{Sungsoo Hong}, \bibinfo{person}{Minhyang Suh}, \bibinfo{person}{Tae~Soo Kim}, \bibinfo{person}{Irina Smoke}, \bibinfo{person}{Sangwha Sien}, \bibinfo{person}{Janet Ng}, \bibinfo{person}{Mark Zachry}, {and} \bibinfo{person}{Juho Kim}.} \bibinfo{year}{2019}\natexlab{b}.
\newblock \showarticletitle{Design for collaborative information-seeking: Understanding user challenges and deploying collaborative dynamic queries}.
\newblock \bibinfo{journal}{\emph{Proceedings of the ACM on Human-Computer Interaction}} \bibinfo{volume}{3}, \bibinfo{number}{CSCW} (\bibinfo{year}{2019}), \bibinfo{pages}{1--24}.
\newblock


\bibitem[Hong et~al\mbox{.}(2020)]%
        {hong2020human}
\bibfield{author}{\bibinfo{person}{Sungsoo~Ray Hong}, \bibinfo{person}{Jessica Hullman}, {and} \bibinfo{person}{Enrico Bertini}.} \bibinfo{year}{2020}\natexlab{}.
\newblock \showarticletitle{Human factors in model interpretability: Industry practices, challenges, and needs}.
\newblock \bibinfo{journal}{\emph{Proceedings of the ACM on Human-Computer Interaction}} \bibinfo{volume}{4}, \bibinfo{number}{CSCW1} (\bibinfo{year}{2020}), \bibinfo{pages}{1--26}.
\newblock


\bibitem[Hong et~al\mbox{.}(2019a)]%
        {hong2019disseminating}
\bibfield{author}{\bibinfo{person}{Sungsoo~Ray Hong}, \bibinfo{person}{Jorge~Piazentin Ono}, \bibinfo{person}{Juliana Freire}, {and} \bibinfo{person}{Enrico Bertini}.} \bibinfo{year}{2019}\natexlab{a}.
\newblock \showarticletitle{Disseminating Machine Learning to domain experts: Understanding challenges and opportunities in supporting a model building process}. In \bibinfo{booktitle}{\emph{CHI 2019 Workshop, Emerging Perspectives in Human-Centered Machine Learning. ACM}}.
\newblock


\bibitem[Hong et~al\mbox{.}(2018b)]%
        {collab1}
\bibfield{author}{\bibinfo{person}{Sungsoo~(Ray) Hong}, \bibinfo{person}{Minhyang~(Mia) Suh}, \bibinfo{person}{Nathalie Henry~Riche}, \bibinfo{person}{Jooyoung Lee}, \bibinfo{person}{Juho Kim}, {and} \bibinfo{person}{Mark Zachry}.} \bibinfo{year}{2018}\natexlab{b}.
\newblock \showarticletitle{Collaborative Dynamic Queries: Supporting Distributed Small Group Decision-Making}. In \bibinfo{booktitle}{\emph{Proceedings of the 2018 CHI Conference on Human Factors in Computing Systems}} (Montreal QC, Canada) \emph{(\bibinfo{series}{CHI '18})}. \bibinfo{publisher}{Association for Computing Machinery}, \bibinfo{address}{New York, NY, USA}, \bibinfo{pages}{1–12}.
\newblock
\showISBNx{9781450356206}
\urldef\tempurl%
\url{https://doi.org/10.1145/3173574.3173640}
\showDOI{\tempurl}


\bibitem[Hong et~al\mbox{.}(2019c)]%
        {10.1145/3359208}
\bibfield{author}{\bibinfo{person}{Sungsoo~(Ray) Hong}, \bibinfo{person}{Minhyang~(Mia) Suh}, \bibinfo{person}{Tae~Soo Kim}, \bibinfo{person}{Irina Smoke}, \bibinfo{person}{Sangwha Sien}, \bibinfo{person}{Janet Ng}, \bibinfo{person}{Mark Zachry}, {and} \bibinfo{person}{Juho Kim}.} \bibinfo{year}{2019}\natexlab{c}.
\newblock \showarticletitle{Design for Collaborative Information-Seeking: Understanding User Challenges and Deploying Collaborative Dynamic Queries}.
\newblock \bibinfo{journal}{\emph{Proc. ACM Hum.-Comput. Interact.}} \bibinfo{volume}{3}, \bibinfo{number}{CSCW}, Article \bibinfo{articleno}{106} (\bibinfo{date}{nov} \bibinfo{year}{2019}), \bibinfo{numpages}{24}~pages.
\newblock
\urldef\tempurl%
\url{https://doi.org/10.1145/3359208}
\showDOI{\tempurl}


\bibitem[Hughes et~al\mbox{.}(2014)]%
        {hughespeterson2014}
\bibfield{author}{\bibinfo{person}{Amanda~L. Hughes}, \bibinfo{person}{Steve Peterson}, {and} \bibinfo{person}{Leysia Palen}.} \bibinfo{year}{2014}\natexlab{}.
\newblock \bibinfo{booktitle}{\emph{Social media in emergency management}}.
\newblock 349--392 pages.
\newblock


\bibitem[Imran et~al\mbox{.}(2022)]%
        {imran2022ai}
\bibfield{author}{\bibinfo{person}{Muhammad Imran}, \bibinfo{person}{Umair Qazi}, \bibinfo{person}{Ferda Ofli}, \bibinfo{person}{Steve Peterson}, {and} \bibinfo{person}{Firoj Alam}.} \bibinfo{year}{2022}\natexlab{}.
\newblock \showarticletitle{Ai for disaster rapid damage assessment from microblogs}. In \bibinfo{booktitle}{\emph{Proceedings of the AAAI Conference on Artificial Intelligence}}, Vol.~\bibinfo{volume}{36}. \bibinfo{pages}{12517--12523}.
\newblock


\bibitem[Inc.(2023a)]%
        {chatgpt}
\bibfield{author}{\bibinfo{person}{OpenAI Inc.}} \bibinfo{year}{2023}\natexlab{a}.
\newblock \bibinfo{title}{ChatGPT}.
\newblock
\newblock
\urldef\tempurl%
\url{https://chat.openai.com/}
\showURL{%
\tempurl}
\newblock
\shownote{Accessed: 2023-09-20}.


\bibitem[Inc.(2023b)]%
        {twitter}
\bibfield{author}{\bibinfo{person}{Twitter Inc.}} \bibinfo{year}{2023}\natexlab{b}.
\newblock \bibinfo{title}{Twitter}.
\newblock
\newblock
\urldef\tempurl%
\url{https://twitter.com/}
\showURL{%
\tempurl}
\newblock
\shownote{Accessed: 2023-09-13}.


\bibitem[Kairam and Heer(2016)]%
        {10.1145/2818048.2820016}
\bibfield{author}{\bibinfo{person}{Sanjay Kairam} {and} \bibinfo{person}{Jeffrey Heer}.} \bibinfo{year}{2016}\natexlab{}.
\newblock \showarticletitle{Parting Crowds: Characterizing Divergent Interpretations in Crowdsourced Annotation Tasks}. In \bibinfo{booktitle}{\emph{Proceedings of the 19th ACM Conference on Computer-Supported Cooperative Work \& Social Computing}} (San Francisco, California, USA) \emph{(\bibinfo{series}{CSCW '16})}. \bibinfo{publisher}{Association for Computing Machinery}, \bibinfo{address}{New York, NY, USA}, \bibinfo{pages}{1637–1648}.
\newblock
\showISBNx{9781450335928}
\urldef\tempurl%
\url{https://doi.org/10.1145/2818048.2820016}
\showDOI{\tempurl}


\bibitem[Kapania et~al\mbox{.}(2023)]%
        {10.1145/3544548.3580645}
\bibfield{author}{\bibinfo{person}{Shivani Kapania}, \bibinfo{person}{Alex~S Taylor}, {and} \bibinfo{person}{Ding Wang}.} \bibinfo{year}{2023}\natexlab{}.
\newblock \showarticletitle{A Hunt for the Snark: Annotator Diversity in Data Practices}. In \bibinfo{booktitle}{\emph{Proceedings of the 2023 CHI Conference on Human Factors in Computing Systems}} (Hamburg, Germany) \emph{(\bibinfo{series}{CHI '23})}. \bibinfo{publisher}{Association for Computing Machinery}, \bibinfo{address}{New York, NY, USA}, Article \bibinfo{articleno}{133}, \bibinfo{numpages}{15}~pages.
\newblock
\showISBNx{9781450394215}
\urldef\tempurl%
\url{https://doi.org/10.1145/3544548.3580645}
\showDOI{\tempurl}


\bibitem[Khan et~al\mbox{.}(2022)]%
        {khan2022transformers}
\bibfield{author}{\bibinfo{person}{Salman Khan}, \bibinfo{person}{Muzammal Naseer}, \bibinfo{person}{Munawar Hayat}, \bibinfo{person}{Syed~Waqas Zamir}, \bibinfo{person}{Fahad~Shahbaz Khan}, {and} \bibinfo{person}{Mubarak Shah}.} \bibinfo{year}{2022}\natexlab{}.
\newblock \showarticletitle{Transformers in vision: A survey}.
\newblock \bibinfo{journal}{\emph{ACM computing surveys (CSUR)}} \bibinfo{volume}{54}, \bibinfo{number}{10s} (\bibinfo{year}{2022}), \bibinfo{pages}{1--41}.
\newblock


\bibitem[Lazar et~al\mbox{.}(2010)]%
        {10.5555/1841406}
\bibfield{author}{\bibinfo{person}{Jonathan Lazar}, \bibinfo{person}{Jinjuan~Heidi Feng}, {and} \bibinfo{person}{Harry Hochheiser}.} \bibinfo{year}{2010}\natexlab{}.
\newblock \bibinfo{booktitle}{\emph{Research Methods in Human-Computer Interaction}}.
\newblock \bibinfo{publisher}{Wiley Publishing}.
\newblock
\showISBNx{0470723378}


\bibitem[Liu et~al\mbox{.}(2019)]%
        {liu2019deep}
\bibfield{author}{\bibinfo{person}{Zimo Liu}, \bibinfo{person}{Jingya Wang}, \bibinfo{person}{Shaogang Gong}, \bibinfo{person}{Huchuan Lu}, {and} \bibinfo{person}{Dacheng Tao}.} \bibinfo{year}{2019}\natexlab{}.
\newblock \showarticletitle{{Deep reinforcement active learning for human-in-the-loop person re-identification}}. In \bibinfo{booktitle}{\emph{Proceedings of the IEEE/CVF international conference on computer vision}}. \bibinfo{pages}{6122--6131}.
\newblock


\bibitem[Lutnick et~al\mbox{.}(2019)]%
        {lutnick2019integrated}
\bibfield{author}{\bibinfo{person}{Brendon Lutnick}, \bibinfo{person}{Brandon Ginley}, \bibinfo{person}{Darshana Govind}, \bibinfo{person}{Sean~D McGarry}, \bibinfo{person}{Peter~S LaViolette}, \bibinfo{person}{Rabi Yacoub}, \bibinfo{person}{Sanjay Jain}, \bibinfo{person}{John~E Tomaszewski}, \bibinfo{person}{Kuang-Yu Jen}, {and} \bibinfo{person}{Pinaki Sarder}.} \bibinfo{year}{2019}\natexlab{}.
\newblock \showarticletitle{{An integrated iterative annotation technique for easing neural network training in medical image analysis}}.
\newblock \bibinfo{journal}{\emph{Nature machine intelligence}} \bibinfo{volume}{1}, \bibinfo{number}{2} (\bibinfo{year}{2019}), \bibinfo{pages}{112--119}.
\newblock


\bibitem[Madono et~al\mbox{.}(2020)]%
        {madono2020efficient}
\bibfield{author}{\bibinfo{person}{Koki Madono}, \bibinfo{person}{Teppei Nakano}, \bibinfo{person}{Tetsunori Kobayashi}, {and} \bibinfo{person}{Tetsuji Ogawa}.} \bibinfo{year}{2020}\natexlab{}.
\newblock \showarticletitle{{Efficient human-in-the-loop object detection using bi-directional deep sort and annotation-free segment identification}}. In \bibinfo{booktitle}{\emph{2020 Asia-Pacific Signal and Information Processing Association Annual Summit and Conference (APSIPA ASC)}}. IEEE, \bibinfo{pages}{1226--1233}.
\newblock


\bibitem[Mengle and Goharian(2009)]%
        {mengle2009ambiguity}
\bibfield{author}{\bibinfo{person}{Saket~SR Mengle} {and} \bibinfo{person}{Nazli Goharian}.} \bibinfo{year}{2009}\natexlab{}.
\newblock \showarticletitle{Ambiguity measure feature-selection algorithm}.
\newblock \bibinfo{journal}{\emph{Journal of the American Society for Information Science and Technology}} \bibinfo{volume}{60}, \bibinfo{number}{5} (\bibinfo{year}{2009}), \bibinfo{pages}{1037--1050}.
\newblock


\bibitem[Pandey et~al\mbox{.}(2022)]%
        {pandey2022modeling}
\bibfield{author}{\bibinfo{person}{Rahul Pandey}, \bibinfo{person}{Hemant Purohit}, \bibinfo{person}{Carlos Castillo}, {and} \bibinfo{person}{Valerie~L Shalin}.} \bibinfo{year}{2022}\natexlab{}.
\newblock \showarticletitle{Modeling and mitigating human annotation errors to design efficient stream processing systems with human-in-the-loop machine learning}.
\newblock \bibinfo{journal}{\emph{International Journal of Human-Computer Studies}}  \bibinfo{volume}{160} (\bibinfo{year}{2022}), \bibinfo{pages}{102772}.
\newblock


\bibitem[Peterson et~al\mbox{.}(2019)]%
        {peterson2019}
\bibfield{author}{\bibinfo{person}{Steve Peterson}, \bibinfo{person}{Keri Stephens}, \bibinfo{person}{Amanda Hughes}, {and} \bibinfo{person}{Hemant Purohit}.} \bibinfo{year}{2019}\natexlab{}.
\newblock \showarticletitle{When Official Systems Overload: A Framework for Finding Social Media Calls for Help during Evacuations}.
\newblock  (\bibinfo{year}{2019}), \bibinfo{pages}{867--875}.
\newblock


\bibitem[Robson et~al\mbox{.}(2020)]%
        {Robson_Searston_Edmond_McCarthy_Tangen_2020}
\bibfield{author}{\bibinfo{person}{Samuel~G. Robson}, \bibinfo{person}{Rachel~A. Searston}, \bibinfo{person}{Gary Edmond}, \bibinfo{person}{Duncan~J. McCarthy}, {and} \bibinfo{person}{Jason~M. Tangen}.} \bibinfo{year}{2020}\natexlab{}.
\newblock \showarticletitle{An expert–novice comparison of feature choice}.
\newblock \bibinfo{journal}{\emph{Applied Cognitive Psychology}} \bibinfo{volume}{34}, \bibinfo{number}{5} (\bibinfo{date}{Sept.} \bibinfo{year}{2020}), \bibinfo{pages}{984–995}.
\newblock
\showISSN{0888-4080, 1099-0720}
\urldef\tempurl%
\url{https://doi.org/10.1002/acp.3676}
\showDOI{\tempurl}


\bibitem[Russell and Chi(2014)]%
        {Russell2014LookingBR}
\bibfield{author}{\bibinfo{person}{Daniel~M. Russell} {and} \bibinfo{person}{Ed~H. Chi}.} \bibinfo{year}{2014}\natexlab{}.
\newblock \showarticletitle{Looking Back: Retrospective Study Methods for HCI}. In \bibinfo{booktitle}{\emph{Ways of Knowing in HCI}}.
\newblock
\urldef\tempurl%
\url{https://api.semanticscholar.org/CorpusID:2470741}
\showURL{%
\tempurl}


\bibitem[Russo et~al\mbox{.}(2021)]%
        {RUSSO2021117695}
\bibfield{author}{\bibinfo{person}{Stefania Russo}, \bibinfo{person}{Michael~D. Besmer}, \bibinfo{person}{Frank Blumensaat}, \bibinfo{person}{Damien Bouffard}, \bibinfo{person}{Andy Disch}, \bibinfo{person}{Frederik Hammes}, \bibinfo{person}{Angelika Hess}, \bibinfo{person}{Moritz Lürig}, \bibinfo{person}{Blake Matthews}, \bibinfo{person}{Camille Minaudo}, \bibinfo{person}{Eberhard Morgenroth}, \bibinfo{person}{Viet Tran-Khac}, {and} \bibinfo{person}{Kris Villez}.} \bibinfo{year}{2021}\natexlab{}.
\newblock \showarticletitle{The value of human data annotation for machine learning based anomaly detection in environmental systems}.
\newblock \bibinfo{journal}{\emph{Water Research}}  \bibinfo{volume}{206} (\bibinfo{year}{2021}), \bibinfo{pages}{117695}.
\newblock
\showISSN{0043-1354}
\urldef\tempurl%
\url{https://doi.org/10.1016/j.watres.2021.117695}
\showDOI{\tempurl}


\bibitem[Santos et~al\mbox{.}(2019)]%
        {santos2019visus}
\bibfield{author}{\bibinfo{person}{A{\'e}cio Santos}, \bibinfo{person}{Sonia Castelo}, \bibinfo{person}{Cristian Felix}, \bibinfo{person}{Jorge~Piazentin Ono}, \bibinfo{person}{Bowen Yu}, \bibinfo{person}{Sungsoo~Ray Hong}, \bibinfo{person}{Cl{\'a}udio~T Silva}, \bibinfo{person}{Enrico Bertini}, {and} \bibinfo{person}{Juliana Freire}.} \bibinfo{year}{2019}\natexlab{}.
\newblock \showarticletitle{Visus: An interactive system for automatic machine learning model building and curation}. In \bibinfo{booktitle}{\emph{Proceedings of the Workshop on Human-In-the-Loop Data Analytics}}. \bibinfo{pages}{1--7}.
\newblock


\bibitem[Senarath et~al\mbox{.}(2022)]%
        {senarath2022citizen}
\bibfield{author}{\bibinfo{person}{Yasas Senarath}, \bibinfo{person}{Rahul Pandey}, \bibinfo{person}{Steve Peterson}, {and} \bibinfo{person}{Hemant Purohit}.} \bibinfo{year}{2022}\natexlab{}.
\newblock \showarticletitle{Citizen-Helper System for Human-Centered AI Use in Disaster Management}.
\newblock In \bibinfo{booktitle}{\emph{International Handbook of Disaster Research}}. \bibinfo{publisher}{Springer}, \bibinfo{pages}{1--21}.
\newblock


\bibitem[Senarath et~al\mbox{.}(2021)]%
        {senarath2021mining}
\bibfield{author}{\bibinfo{person}{Yasas Senarath}, \bibinfo{person}{Steve Peterson}, \bibinfo{person}{Hemant Purohit}, \bibinfo{person}{Amanda~L Hughes}, {and} \bibinfo{person}{Keri~K Stephens}.} \bibinfo{year}{2021}\natexlab{}.
\newblock \showarticletitle{Mining risk behaviors from social media for pandemic crisis preparedness and response}. In \bibinfo{booktitle}{\emph{Proceedings of the 2021 international conference on social computing, behavioral-cultural modeling \& prediction and behavior representation in modeling and simulation}}.
\newblock


\bibitem[Serrano et~al\mbox{.}(2013)]%
        {serrano2013cloud}
\bibfield{author}{\bibinfo{person}{Mart{\'\i}n Serrano}, \bibinfo{person}{Lei Shi}, \bibinfo{person}{M{\'\i}che{\'a}l~{\'O} Foghl{\'u}}, {and} \bibinfo{person}{William Donnelly}.} \bibinfo{year}{2013}\natexlab{}.
\newblock \showarticletitle{Cloud services composition support by using semantic annotation and linked data}. In \bibinfo{booktitle}{\emph{Knowledge Discovery, Knowledge Engineering and Knowledge Management: Third International Joint Conference, IC3K 2011, Paris, France, October 26-29, 2011. Revised Selected Papers 3}}. Springer, \bibinfo{pages}{278--293}.
\newblock


\bibitem[Service(2023)]%
        {Ian}
\bibfield{author}{\bibinfo{person}{National~Weather Service}.} \bibinfo{year}{2023}\natexlab{}.
\newblock \bibinfo{title}{Hurricane IAN}.
\newblock
\newblock
\urldef\tempurl%
\url{https://www.weather.gov/mhx/HurricaneIan093022}
\showURL{%
\tempurl}
\newblock
\shownote{Accessed: 2023-09-13}.


\bibitem[Shahi and Majchrzak(2021)]%
        {shahi2021amused}
\bibfield{author}{\bibinfo{person}{Gautam~Kishore Shahi} {and} \bibinfo{person}{Tim~A Majchrzak}.} \bibinfo{year}{2021}\natexlab{}.
\newblock \showarticletitle{Amused: an annotation framework of multimodal social media data}. In \bibinfo{booktitle}{\emph{International Conference on Intelligent Technologies and Applications}}. Springer, \bibinfo{pages}{287--299}.
\newblock


\bibitem[Shneiderman and Kang(2000)]%
        {shneiderman2000direct}
\bibfield{author}{\bibinfo{person}{Ben Shneiderman} {and} \bibinfo{person}{Hyunmo Kang}.} \bibinfo{year}{2000}\natexlab{}.
\newblock \showarticletitle{Direct annotation: A drag-and-drop strategy for labeling photos}. In \bibinfo{booktitle}{\emph{2000 IEEE Conference on Information Visualization. An International Conference on Computer Visualization and Graphics}}. IEEE, \bibinfo{pages}{88--95}.
\newblock


\bibitem[Stureborg et~al\mbox{.}(2023)]%
        {Stureborg_2023}
\bibfield{author}{\bibinfo{person}{Rickard Stureborg}, \bibinfo{person}{Bhuwan Dhingra}, {and} \bibinfo{person}{Jun Yang}.} \bibinfo{year}{2023}\natexlab{}.
\newblock \showarticletitle{Interface Design for Crowdsourcing Hierarchical Multi-Label Text Annotations}. In \bibinfo{booktitle}{\emph{Proceedings of the 2023 {CHI} Conference on Human Factors in Computing Systems}}. \bibinfo{publisher}{{ACM}}.
\newblock
\urldef\tempurl%
\url{https://doi.org/10.1145/3544548.3581431}
\showDOI{\tempurl}


\bibitem[Sun et~al\mbox{.}(2023)]%
        {sun2023designing}
\bibfield{author}{\bibinfo{person}{Tong~Steven Sun}, \bibinfo{person}{Yuyang Gao}, \bibinfo{person}{Shubham Khaladkar}, \bibinfo{person}{Sijia Liu}, \bibinfo{person}{Liang Zhao}, \bibinfo{person}{Young-Ho Kim}, {and} \bibinfo{person}{Sungsoo~Ray Hong}.} \bibinfo{year}{2023}\natexlab{}.
\newblock \showarticletitle{Designing a Direct Feedback Loop between Humans and Convolutional Neural Networks through Local Explanations}.
\newblock \bibinfo{journal}{\emph{Proceedings of the ACM on Human-Computer Interaction}} \bibinfo{volume}{7}, \bibinfo{number}{CSCW2} (\bibinfo{year}{2023}), \bibinfo{pages}{1--32}.
\newblock


\bibitem[Sutcliffe(2005)]%
        {10.1145/1082983.1083119}
\bibfield{author}{\bibinfo{person}{Alistair Sutcliffe}.} \bibinfo{year}{2005}\natexlab{}.
\newblock \showarticletitle{Applying small group theory to analysis and design of CSCW systems}.
\newblock \bibinfo{journal}{\emph{SIGSOFT Softw. Eng. Notes}} \bibinfo{volume}{30}, \bibinfo{number}{4} (\bibinfo{date}{may} \bibinfo{year}{2005}), \bibinfo{pages}{1–6}.
\newblock
\showISSN{0163-5948}
\urldef\tempurl%
\url{https://doi.org/10.1145/1082983.1083119}
\showDOI{\tempurl}


\bibitem[Upchurch et~al\mbox{.}(2016)]%
        {Upchurch_Sedra_Mullen_Hirsh_Bala_2016}
\bibfield{author}{\bibinfo{person}{Paul Upchurch}, \bibinfo{person}{Daniel Sedra}, \bibinfo{person}{Andrew Mullen}, \bibinfo{person}{Haym Hirsh}, {and} \bibinfo{person}{Kavita Bala}.} \bibinfo{year}{2016}\natexlab{}.
\newblock \showarticletitle{Interactive Consensus Agreement Games for Labeling Images}.
\newblock \bibinfo{journal}{\emph{Proceedings of the AAAI Conference on Human Computation and Crowdsourcing}}  \bibinfo{volume}{4} (\bibinfo{date}{Sept.} \bibinfo{year}{2016}), \bibinfo{pages}{239–248}.
\newblock
\showISSN{2769-1349, 2769-1330}
\urldef\tempurl%
\url{https://doi.org/10.1609/hcomp.v4i1.13293}
\showDOI{\tempurl}


\bibitem[Wang et~al\mbox{.}(2022)]%
        {wang2022ai}
\bibfield{author}{\bibinfo{person}{Ding Wang}, \bibinfo{person}{Shantanu Prabhat}, {and} \bibinfo{person}{Nithya Sambasivan}.} \bibinfo{year}{2022}\natexlab{}.
\newblock \bibinfo{title}{Whose AI Dream? In search of the aspiration in data annotation}.
\newblock
\newblock
\showeprint[arxiv]{2203.10748}~[cs.HC]


\bibitem[Wilkins et~al\mbox{.}(2022)]%
        {10.1145/3512940}
\bibfield{author}{\bibinfo{person}{Denise~J. Wilkins}, \bibinfo{person}{Srihari Hulikal~Muralidhar}, \bibinfo{person}{Max Meijer}, \bibinfo{person}{Laura Lascau}, {and} \bibinfo{person}{Si\^{a}n Lindley}.} \bibinfo{year}{2022}\natexlab{}.
\newblock \showarticletitle{Gigified Knowledge Work: Understanding Knowledge Gaps When Knowledge Work and On-Demand Work Intersect}.
\newblock \bibinfo{journal}{\emph{Proc. ACM Hum.-Comput. Interact.}} \bibinfo{volume}{6}, \bibinfo{number}{CSCW1}, Article \bibinfo{articleno}{93} (\bibinfo{date}{apr} \bibinfo{year}{2022}), \bibinfo{numpages}{27}~pages.
\newblock
\urldef\tempurl%
\url{https://doi.org/10.1145/3512940}
\showDOI{\tempurl}


\bibitem[Yan et~al\mbox{.}(2022)]%
        {yan2022flatmagic}
\bibfield{author}{\bibinfo{person}{Chuan Yan}, \bibinfo{person}{John Joon~Young Chung}, \bibinfo{person}{Yoon Kiheon}, \bibinfo{person}{Yotam Gingold}, \bibinfo{person}{Eytan Adar}, {and} \bibinfo{person}{Sungsoo~Ray Hong}.} \bibinfo{year}{2022}\natexlab{}.
\newblock \showarticletitle{FlatMagic: Improving flat colorization through AI-driven design for digital comic professionals}. In \bibinfo{booktitle}{\emph{Proceedings of the 2022 CHI Conference on Human Factors in Computing Systems}}. \bibinfo{pages}{1--17}.
\newblock


\bibitem[Zhou et~al\mbox{.}(2023)]%
        {zhou-etal-2023-context}
\bibfield{author}{\bibinfo{person}{Wenxuan Zhou}, \bibinfo{person}{Sheng Zhang}, \bibinfo{person}{Hoifung Poon}, {and} \bibinfo{person}{Muhao Chen}.} \bibinfo{year}{2023}\natexlab{}.
\newblock \showarticletitle{Context-faithful Prompting for Large Language Models}. In \bibinfo{booktitle}{\emph{Findings of the Association for Computational Linguistics: EMNLP 2023}}, \bibfield{editor}{\bibinfo{person}{Houda Bouamor}, \bibinfo{person}{Juan Pino}, {and} \bibinfo{person}{Kalika Bali}} (Eds.). \bibinfo{publisher}{Association for Computational Linguistics}, \bibinfo{address}{Singapore}, \bibinfo{pages}{14544--14556}.
\newblock
\urldef\tempurl%
\url{https://doi.org/10.18653/v1/2023.findings-emnlp.968}
\showDOI{\tempurl}


\end{thebibliography}
